\documentclass[onecolumn,preprintnumbers,pre]{revtex4}
\usepackage{graphicx, subfigure}
\usepackage{amssymb, amsmath,amssymb,amsfonts}
\usepackage{amsthm,mathrsfs,amsopn}
\usepackage{dcolumn}
\usepackage{bm}
\usepackage{xcolor}
\usepackage[utf8]{inputenc}
\usepackage{mathtools}
\usepackage[english]{babel}

\usepackage{graphicx} % For rotation

\definecolor{internationalkleinblue}{rgb}{0.0, 0.18, 0.65}
\definecolor{applegreen}{rgb}{0.55, 0.71, 0.0}
\definecolor{darkviolet}{rgb}{0.58, 0.0, 0.83}
\definecolor{darkpink}{rgb}{0.91, 0.33, 0.5}

\usepackage[colorlinks=true, allcolors=blue]{hyperref}

\theoremstyle{plain} %% This is the default

\newtheorem{theorem}{Theorem}

\newtheorem{remark}[theorem]{Remark}

\begin{document}

\title{Impact of directionality on the emergence of Turing patterns on $m$-directed higher-order structures}

\author{Marie Dorchain$^{1,*}$}
\author{Wilfried Segnou$^{1,*}$}
\author{Riccardo Muolo$^{2}$}
\author{Timoteo Carletti$^{1,3}$}
\affiliation{$^1$ Department of mathematics and Namur Institute for Complex Systems, naXys, University of Namur, Belgium, timoteo.carletti@unamur.be}
\address{$^{2}$Department of Systems and Control Engineering, Tokyo Institute of Technology, Japan}
\affiliation{$^*$ Those authors equally contributed to the work}
\affiliation{$^3$ timoteo.carletti@unamur.be}

\begin{abstract}
We hereby develop the theory of Turing instability for reaction–diffusion systems defined on $m$-directed hypergraphs, the latter being generalization of hypergraphs where nodes forming hyperedges can be shared into two disjoint sets, the head nodes and the tail nodes. This framework encodes thus for a privileged direction for the reaction to occur: the joint action of tail nodes is a driver for the reaction involving head nodes. It thus results a natural generalization of directed networks. Based on a linear stability analysis we have shown the existence of two Laplace matrices, allowing to analytically prove that Turing patterns, stationary or wave-like, emerges for a much broader set of parameters in the $m$-directed setting. In particular directionality promotes Turing instability, otherwise absent in the symmetric case. Analytical results are compared to simulations performed by using the Brusselator model defined on a $m$-directed $d$-hyperring as well as on a $m$-directed random hypergraph.
\end{abstract}
\maketitle

\section{Introduction}
\label{sec:intro}

A blossoming of regular spatio-temporal complex patterns surrounds us, they are the signature of self-organized processes where ordered structures emerge from disordered ones because of the intricate web of nonlinear interactions existing between the basic units by which the system under scrutiny is made of~\cite{PrigogineNicolis1967,Murray2001,PSV2010}. In many applications, the interaction between microscopic units can be modeled by means of reaction-diffusion equations that govern the deterministic evolution of species concentrations across time and space. One of the most elegant and widespread theory applied to explain the emergence of self-organized patterns in this framework is the one due to Alan Turing, who proposed a mechanism of patterns-formation rooted on a diffusion-driven instability, which now bears his name~\cite{Turing1952}. Turing instability results from the combined action of a (local) reaction and a (long-range) diffusion, involving an activator and an inhibitor species~\cite{GiererMeinhardt}. Let us emphasize that each process considered separately would drive the system to a spatially homogeneous state; there are however conditions on the models and on the interactions such that any heterogeneous, arbitrarily small, perturbation of the homogeneous state is amplified and eventually returns a macroscopic patchy (non-homogeneous) solution, i.e., a Turing patterns. The diffusive terms being the destabilizing factors, the above mechanism is known in the literature as a diffusion-driven instability and its application goes well beyond the original framework of morphogenesis or chemical reaction systems and it nowadays stands for a pillar to explain self-organization in Nature~\cite{Pismen06,PrigogineNicolis1967,Murray2001,PSV2010}. Starting from these premises, scholars have been able to extend the original Turing mechanism to non-autonomous systems, e.g., evolving domains~\cite{CGM1999,PSGPBM2004} or time dependent diffusion and reaction rates~\cite{vangorder}, as well as to discrete substrates, e.g., lattices~\cite{OS1971} or complex networks~\cite{NM2010}, and their generalization, e.g., directed networks~\cite{asllani2014theory}, non-normal networks~\cite{jtb2019}, multiplex networks~\cite{Asllani2014} and, recently, to time varying networks~\cite{PABFC2017,vangorder2,CarlettiFanelli2022}. The interested reader can consult the two recent review papers on Turing instability on continuous domains~\cite{KGMK2021} and networks~\cite{MGNFC2024}.

The extension of Turing instability to complex networks is a natural one, as many real and engineered systems, e.g., the brain and power grids, find a natural embedding in a networked support. Indeed in such framework, the local reactions involve spatially very close species separated from other groups, defining this the nodes of a networks, while the diffusion takes place across paths connecting different zones, i.e., the links of a network. Despite the success of network science~\cite{AB2002,BLMChH2006} in describing many empirical systems~\cite{newmanbook,barabasibook,Latorabook}, mainly because networks constitute an abstract framework, where pairwise interactions (edges) among generic agents (nodes) are encoded, scholars recently realized that several examples exist for which the pairwise approximation holds true only as a first order step and the group action is the real driver of the dynamics~\cite{LRS2019,petri2014homological,LEFGHVD2016,estradaJTB,Abrams1983,GBMSA,patania2017shape,battiston2020networks}. Higher-orders structures have been brought to the fore to overcome this limitation, indeed they allow us to capture the many-body interactions among the units, the main examples being simplicial complexes~\cite{DVVM,BC,petri2018simplicial} and hypergraphs~\cite{berge1973graphs,estrada2005complex,GZCN}. Those are generalizations of networks that are attracting a lot of interest from scholars of different fields, for example to study spreading of viruses~\cite{IPBL,de2019social}, random walks~\cite{CarlettiEtAl2020,TTTH2015}, and to the study of synchronization~\cite{Krawiecki2014,MKJ2020} or diffusion~\cite{ATM2020}.
 
 Turing theory has been extended to hypergraphs~\cite{CarlettiFanelliNicoletti2020} by considering a sort of hyperedge mean-field, being the nodes inside a hyperedge equivalent with each other, the impact of any of them onto another node, still in the same hyperedge, is modulated by a function of the size of the hyperedge. In this way, authors have been able to map the system defined on the hypergraph into a system defined on a suitable weighted network. Later, a general theory of Turing patterns on higher-order structures has been developed in~\cite{muolo2023turinghh}, where hypergraphs structures have been encoded into symmetric adjacency tensors to encompass for many-body interactions. The aim of this paper is to take one step further by proposing an extension of Turing instability on directed higher-order topologies by adapting the formalism developed in the context of synchronization on higher-order structures in~\cite{gallo2022synchronization}. 
 
Indeed in some relevant research cases interactions are not symmetric: the action of agent $A$ upon $B$ is not the same of $B$ against $A$. In the pairwise setting this implies to deal with directed network, while in the case of many-body interactions one can consider $m$-directed higher-order structures~\cite{gallo2022synchronization}. In this present work we develop further the latter and in particular we clarify the role of the directionality, $m$, with respect to~\cite{gallo2022synchronization} where the framework has been introduced to study synchronization on $1$-directed higher-order structures. Directed hypergraphs have been studied in the past mainly in operational research and in computer science, but without developing a full algebraic formalism as the one hereby presented, see, e.g., the concepts of B-arc and B-hypergraph introduced by~\cite{GLPN1993}, the structures proposed in~\cite{AUSIELLO2017293} in order to describe and discuss optimization problems, or the work~\cite{DucournauBretto2014} devoted to computer vision and images segmentation. A step forward in the direction of building an algebraic formalism has been done in~\cite{JOST2019870} where authors defined the normalized Laplace matrix for chemical reaction networks, that corresponds to an {\em oriented hypergraphs} where nodes inside an hyperedge, encoding for a chemical reaction, have been partitioned into two (possibly not disjoint) classes corresponding to chemical elements taking part to the reaction. A similar approach has been proposed in~\cite{FaccinPRE2022} where tail and head incidence matrices have been defined to perform nodes ranking and community detection on directed hypergraphs. More recently, directed uniform hypergraphs have been introduced to define pinning control strategy to achieve synchronization~\cite{shi2023synchronization}. However let us observe that the proposed definition of many-body interactions is based on the existence of pairwise interactions inside each hyperedge, making it difficult to disentangle the role of pairwise from higher-order interactions. Higher-order interactions denote the fact that the state of a node is influenced by a group of nodes, meaning that it receives information directly from these nodes, without passing through intermediary nodes, as suggested in~\cite{shi2023synchronization}. The formalism we hereby develop is close to the one proposed for directed uniform hypergraphs~\cite{kezan2023synchronization} also to study pinning control synchronization; let us observe that the work~\cite{kezan2023synchronization} still deals with symmetric Laplace matrices, while in the present work we will show that two Laplace matrices spontaneously emerge from the theory, one of which being asymmetric, it will thus carry the whole information about the directionality.

The onset of Turing instability ultimately relies on the study of the spectral properties of a suitable operator built by using the Jacobian of the reaction part and the diffusion term, i.e., the Laplace operator. By assuming the existence of an eigenbasis for the latter, one can compute the dispersion relation, that ultimately determines the onset of the instability as a function of the discrete Laplace spectrum. Already in the case of directed networks, scholars have shown that network asymmetry can trigger the system unstable whereas this is not possible by assuming a symmetric support, seeding thus the emergence of the so-called topology driven patterns~\cite{asllani2014theory}. In the present work we show that a similar result can be obtained in the case of $m$-directed $d$-hypergraphs, a generalization of directed network where, i.e., $d+1$ nodes inside an hyperedge are divided into two groups, $m$ {\em head nodes} and $q=d+1-m$ {\em tail nodes}, the former receiving the action of the latter ones, introducing thus a directionality in the process.

By assuming the presence of a homogeneous solution for the interconnected system and by performing a linear stability analysis, we bring to the fore the existence of two higher-order Laplace matrices, that together with the Jacobian matrices of the reaction and coupling terms are the driver for the onset of Turing instability. Being one of the two higher-order Laplace matrices asymmetric, its eigenvalues are generally complex numbers; we will thus prove the existence of a condition, similar to the one proposed in~\cite{asllani2014theory}, determining the positivity of the dispersion relation. We then present suitable choices of couplings and directed higher-order topologies to analytically determine the conditions for the instability to occur by focusing on the role of the directionality parameter, $m$.

The proposed theory will be complemented by numerical results showing that the size of the head of the hyperedges, i.e., $m$, can favor the emergence of patterns. Interestingly enough we will show the existence of cases for which a symmetric hypergraph can never develop a Turing instability while a directed counterpart does. Eventually, we will show the existence of peculiar choices of parameters for which the opposite holds true: a strong directionality impedes the emergence of patterns.

The paper is organized as follows.
{In section~\ref{sec:model} we present the general framework of dynamical systems defined on top of $m$-directed $d$-hypergraphs. By studying the stability of a stationary solution via a linearization process, we will bring to the fore the existence of two higher-order Laplace matrices associated to the $m$-directed $d$-hypergraph, one of which is asymmetric. In section~\ref{sec:TP}, we focus on two-dimensional reaction-diffusion systems, as often done in the study of Turing patterns, and we develop the conditions for the onset of Turing instability. Section~\ref{sec:dirtoundir} presents a dedicated study of the impact of the hypergraph directionality on the emergence of patterns by introducing a family of weighted and directed hypergraphs containing as a special case an undirected hypergraph.

\section{The model}
\label{sec:model}

Let us consider a set of $N$ basic interacting units whose internal state can be described by a vector $\vec{x}_i\in\mathbb{R}^n$, for all $i=1,\dots,N$. A many-body interaction, in the specific case of $(d+1)$ interacting units, occurs once the time evolution of the system state of a given unit, say $i$, depends on the system states of other $d$ units and of itself, namely
\begin{equation}
\label{eq:manybody}
\frac{d\vec{x}_i}{dt} \propto  \vec{g}^{(d)}\left(\vec{x}_i,\vec{x}_{j_1},\dots,\vec{x}_{j_d}\right) A^{(d)}_{ij_1\dots j_d}\, ,
\end{equation}
for some indexes $j_1,\dots,j_d\in \{1,\dots, N\}$ denoting the units interacting with the $i$-th one. The function $\vec{g}^{(d)}:\mathbb{R}^{(d+1)n}\rightarrow \mathbb{R}^{n}$ determines the {\em functional form} of the interaction, and $A^{(d)}_{ij_1\dots j_d}$ is a $(d+1)$-tensor encoding the {\em interconnection} existing among the latter, i.e., $A^{(d)}_{ij_1\dots j_d}=1$ being the units $i,j_1,\dots, j_d$ interacting each other, and $0$ otherwise. Let us observe that the latter can be considered to be the {\em hyper adjacency} tensor encoding for a $d$-hyperedge~\footnote{We hereby assume a $d$-hyperedge to contain $(d+1)$ nodes. This is the notation used for simplicial complexes but also by some scholars for hypergraphs.} of a given hypergraphs, indeed $A^{(d)}_{ij_1\dots j_d}=1$ does not imply that subgroups of the same basic units also interact with each other as one should impose in the case of simplicial complexes.

If the system allows for interactions up to $(D+1)$-body then we can generally write
\begin{equation}
\label{eq:manybodyD}
\frac{d\vec{x}_i}{dt} = \vec{g}^{(0)}\left(\vec{x}_i\right)+\sum_{d=1}^D \sigma_d \sum_{j_1,\dots,j_d}\vec{g}^{(d)}\left(\vec{x}_i,\vec{x}_{j_1},\dots,\vec{x}_{j_d}\right) A^{(d)}_{ij_1\dots j_d}\quad\forall i=1,\dots,N\, ,
\end{equation}
where $\vec{g}^{(0)}\left(\vec{x}_i\right)$ describes the evolution of the state isolated from the rest of the system, and we have introduced the scalars $\sigma_d$ to modulate the intensity of the coupling, i.e., we assume to somehow normalize $\vec{g}^{(d)}$.  Observe that the indexes $j_1,\dots,j_d$ in the sum run over the whole set of nodes, i.e., $1,\dots, N$. To lighten the notations, we will not explicitly write in the rest of the work the index ranges once the latter ones will be clear from the context.

In the case the interaction is invariant by permuting the indexes of the tensor, i.e., $A^{(d)}_{\pi (i,j_1,\dots,j_d)}$ keeps the same value for any permutation of the $(d+1)$ indexes, then we are dealing with {\em symmetric} higher-order structures, or symmetric hypergraph. This is the formalism developed in~\cite{muolo2023turinghh} to study Turing patterns on hypergraph and it thus widens the study of patterns emergence beyond the pairwise network case. Let us observe that the same framework has been used to study synchronization in higher-order structures~\cite{gambuzza2021synchronization}.

As already stated, there are interesting cases where interactions are not symmetric: the action of agent $A$ upon $B$ is not the same of $B$ against $A$. In the network case, this assumption returns a directed network, while in the case of many-body interactions we are dealing with $m$-directed $d$-higher-order structures~\cite{gallo2022synchronization}. Let us consider thus $(d+1)$ interacting basic units and assume the $d$ units acting on $\vec{x}_i$ to be split into two groups, a first group composed by $1\leq m\leq d$ agents, to which $i$ belongs, and the $q=d+1-m$ remaining agents. The latter ones act onto $i$ but they do not receive any feedback from nodes in the first group; we will thus claim to deal with $m$-nodes in the {\em head} and $q$-nodes in the {\em tail} of the $d$-hyperedge. Llet us observe that in the literature one could also find the names target and source nodes. In the following we will denote the parameter $m$ as {\em directionality}. Because all nodes in the head interact with each other and nodes in the tail interact interchangeably with nodes in the head, the adjacency tensor encoding for this $d$-hyperedge should satisfy
\begin{equation}
\label{eq:}
A^{(d,m)}_{\pi_1(ii_1,\dots i_{m-1})\pi_2(j_1\dots j_q)}=1\, ,
\end{equation}
for any permutation $\pi_1$ involving the indexes $i,i_1,\dots,i_{m-1}$~\footnote{Let us observe that if the head contains only the node $i$, i.e., $m=1$, then the set of indexes $i,i_1,\dots,i_{m-1}$ reduces to $i$ alone.} and any permutation $\pi_2$ involving the indexes $j_1,\dots,j_{q}$ (remember that $m+q=d+1$)~\footnote{Let us comment on the definition we used. The state variable of a node in the head will evolve because of the influence of the remaining nodes in the head and in the tail. The very nature of the $d$-hyperedge formed by $(d+1)$-node determines thus the evolution of the former, for this reason we state that nodes in the tail interact each other to influence nodes in the head, if this were not the case we could not speak of $(d+1)$-many-body interactions. However this does not imply that the evolution of the state variable of a node in the tail is influenced by other nodes in the tail, they can only work together to act on head nodes.}. Observe that if we are considering $d$-body interactions, the fact that the head (resp. tail) nodes interact with each other does not imply automatically the existence of a $(m-1)$-hyperedge associated to $m$ head nodes (resp. $(q-1)$-hyperedge associated to $q$ tail nodes). We are thus not assuming a closure with respect to the inclusion as in the case of simplicial complexes. To lighten the notations we will not explicitly write in the following the permutations, and thus we will denote the adjacency tensor of a $m$ directed $d$-hyperedge with $A^{(d,m)}_{(ii_1\dots i_{m-1})(j_1\dots j_q)}$, where it is clear that the first group of $m$ indexes denotes the head and the remaining $q$ the tail (see Fig.~\ref{fig:1directed3directed} for some simple examples of $m$-directed $d$-hyperedges).
\begin{figure}[h!]
    \centering
    \includegraphics[width=18cm ]{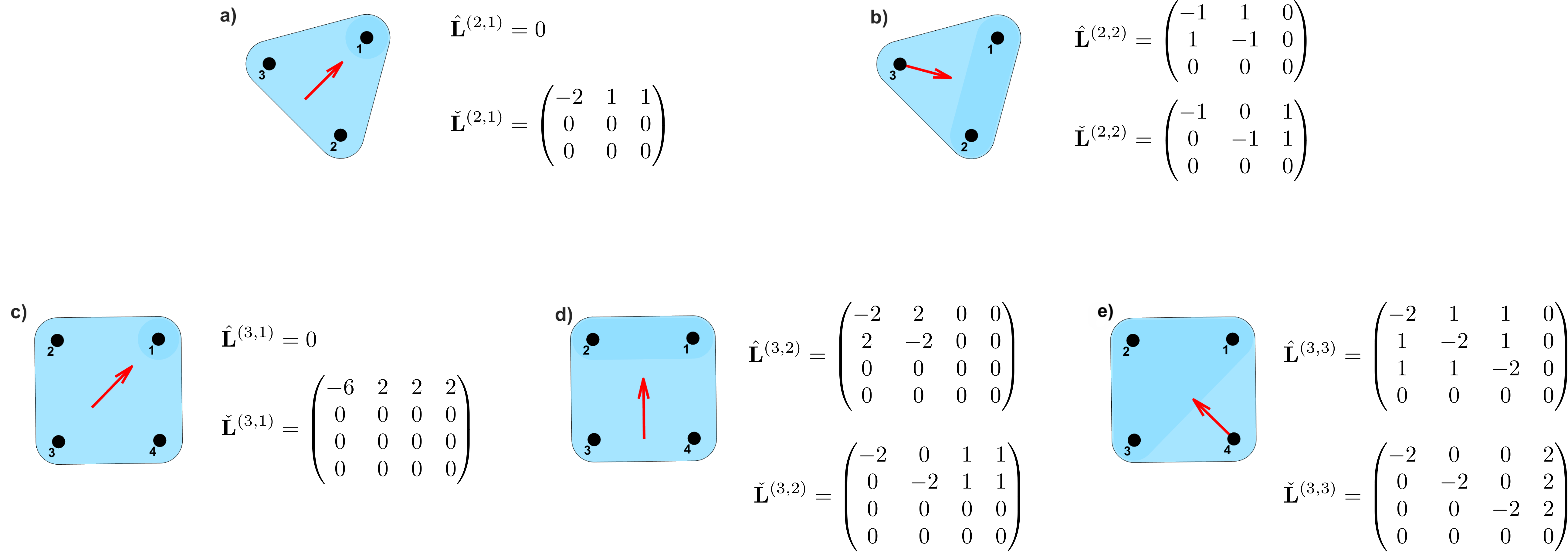}
    \caption{Some examples of $m$-directed $d$-hyperedges and their associated Laplace matrices $\hat{\;\mathbf{L}}^{(d,m)}$ and $\check{\;\mathbf{L}}^{(d,m)}$. In the top panels we schematically show $2$-hyperedges ($1$-directed in panel a) and $2$-directed in panel b), while the bottom panels refer to $3$-hyperedges ($1$-directed in panel c), $2$-directed in panel d) and $3$-directed in panel e)). The hyperedge heads are denoted by a darker blue color and the arrow helps to visualize the directionality. The adjacency tensors are given by $A_{(1)(23)}^{(2,1)}=A_{(1)(32)}^{(2,1)}=1$ (a), $A_{(12)(3)}^{(2,2)}=A_{(21)(3)}^{(2,2)}=1$ (b), $A_{(1)(234)}^{(3,1)}=A_{(1)(243)}^{(3,1)}=A_{(1)(324)}^{(3,1)}=A_{(1)(342)}^{(3,1)}=A_{(1)(423)}^{(3,1)}=A_{(1)(432)}^{(3,1)}=1$ (c), $A_{(12)(34)}^{(3,2)}=A_{(21)(34)}^{(3,2)}=A_{(12)(43)}^{(3,2)}=A_{(21)(43)}^{(3,2)}=1$ (d) and $A_{(123)(4)}^{(3,3)}=A_{(132)(4)}^{(3,3)} =A_{(213)(4)}^{(3,3)}=A_{(231)(4)}^{(3,3)}=A_{(312)(4)}^{(3,3)}=A_{(321)(4)}^{(3,3)}=1$ (e), all the remaining entries vanish. Beside each hyperedge we show the associated matrices $\hat{\;\mathbf{L}}^{(d,m)}$ and $\check{\;\mathbf{L}}^{(d,m)}$.}
    \label{fig:1directed3directed}
\end{figure}

Eventually, a generic system encompassing for directed higher-order interactions can be cast in the following form $\forall i=1,\dots,N$
\begin{equation}
\label{eq:manybodymD}
\frac{d\vec{x}_i}{dt} = \vec{g}^{(0)}\left(\vec{x}_i\right)+\sum_{d=1}^D \sigma_d \sum_{m=1}^d\sum_{\substack{i_1,\dots,i_{m-1}\\ j_1,\dots,j_q}} \vec{g}^{(d,m)}\left(\vec{x}_i,\vec{x}_{i_1},\dots,\vec{x}_{i_{m-1}}\vec{x}_{j_1},\dots,\vec{x}_{j_q}\right) A^{(d,m)}_{(ii_1\dots i_{m-1})(j_1\dots j_q)}\, .
\end{equation}
Here $\vec{g}^{(d,m)}$ is the coupling function, whose notation emphasizes the existence of $m$-head nodes and $q$-tail nodes, where $q=d+1-m$. Moreover, if $m=1$ the sum over the indexes $i_1,\dots,i_{m-1}$ is empty.

Synchronization and Turing instability are both based on a linear stability analysis of system~\eqref{eq:manybodymD} about a synchronous solution, that results to be spatially homogeneous and time-dependent in the former case and stationary in the second one. Let us thus assume there exists such a solution $\vec{x}_i(t)=\vec{x}^*(t)\in\mathbb{R}^n$  for all $i=1,\dots, N$ of the isolated system, namely $\displaystyle \frac{d\vec{x}^*}{dt}=\vec{g}^{(0)}\left(\vec{x}^*\right)$. To ensure that such equilibrium is also a solution of the full system~\eqref{eq:manybodymD}, we have to require the whole term involving the sums to vanish. A straightforward hypothesis widely used in the literature to achieve this goal is to assume the functions $\vec{g}^{(d,m)}$ to be {\em diffusive-like}, namely
\begin{equation}
\label{eq:difflike}
\vec{g}^{(d,m)}\left(\vec{x}_i,\vec{x}_{i_1},\dots,\vec{x}_{i_{m-1}}\vec{x}_{j_1},\dots,\vec{x}_{j_q}\right)=\vec{h}^{(d,m)}\left(\vec{x}_{i_1},\dots,\vec{x}_{i_{m-1}}\vec{x}_{j_1},\dots,\vec{x}_{j_q}\right)-\vec{h}^{(d,m)}\left(\vec{x}_i,\dots,\vec{x}_i\right)\, ,
\end{equation}
for some functions $\vec{h}^{(d,m)}:\mathbb{R}^{nd}\rightarrow\mathbb{R}^{n}$. In this way, we trivially get that $\vec{g}^{(d,m)}(\vec{x},\dots,\vec{x})=0$ for all $d$ and all $\vec{x}\in\mathbb{R}^n$.

The Turing instability relies on the fact that the solution $\vec{x}_i(t)=\vec{x}^*$ for all $i=1,\dots, N$ is stable once $\sigma_d=0$ for all $d=1,\dots, D$, i.e., once the units behave independently from each other, while it becomes unstable whenever some coupling is in action. Any arbitrarily small perturbation added to this reference solution is thus amplified and the system will converge to another state, generically, dependent on nodes index, i.e., there is spatial dependence of the patchy solution. To determine the onset of Turing instability we thus perform a linear stability analysis of Eq.~\eqref{eq:manybodymD}. More precisely we define $\delta\vec{x}_i=\vec{x}_i-\vec{x}^*$, we rewrite the latter equation in terms of $\delta\vec{x}_i$ and we determine its (short) time evolution by retaining only first order terms. By exploiting the diffusive-like assumption, a straightforward computation allows to obtain for all $i=1,\dots,N$
\begin{eqnarray}
\label{eq:manybodymDlin}
\frac{d\delta\vec{x}_i}{dt} = \mathbf{J}^{(0)}\delta\vec{x}_i+\sum_{d=1}^D \sigma_d \sum_{m=1}^d\sum_{\substack{i_1,\dots,i_{m-1} \\j_1,\dots,j_q}}\Big[\mathbf{J}_1 \vec{h}^{(d,m)}\delta x_{i_1}-\mathbf{J}_1 \vec{h}^{(d,m)}\delta x_{i}+\dots+ \mathbf{J}_{m-1} \vec{h}^{(d,m)}\delta x_{i_{m-1}}-\mathbf{J}_{m-1} \vec{h}^{(d,m)}\delta x_{i}+\notag\\+\mathbf{J}_{m} \vec{h}^{(d,m)}\delta x_{j_1}-\mathbf{J}_{m} \vec{h}^{(d,m)}\delta x_{i}+\dots+ \mathbf{J}_{d} \vec{h}^{(d,m)}\delta x_{j_{q}}-\mathbf{J}_{d} \vec{h}^{(d,m)}\delta x_{i}\Big]A^{(d,m)}_{(ii_1\dots i_{m-1})(j_1\dots j_q)}\, ,
\end{eqnarray}
where $\mathbf{J}^{(0)}$ is the Jacobian of $\vec{g}^{(0)}$ evaluated on the reference solution while $\mathbf{J}_{\ell} \vec{h}^{(d,m)}$ is the Jacobian of $\vec{h}^{(d,m)}$ with respect to the $\ell$-th variable evaluated on the reference solution $\vec{x}^*$, i.e., $\mathbf{J}_\ell \vec{h}^{(d,m)}: = \frac{\partial \vec{h}^{(d,m)}}{\partial \vec{x}_\ell} (\vec{x}^*,\dots, \vec{x}^*)$, $\ell=1,\dots,d$.

Let us recall that the $i$-th node belongs to the head of the considered hyperedge, moreover there are two kinds of terms involving the derivatives of $\vec{h}^{(d,m)}$: those for which the derivative is performed with respect to a variable in the head of the hyperedge, and, moreover those associated to variables in the tail. We can thus define two matrices, $\hat{\;\mathbf{L}}^{(d,m)}$ and $\check{\;\mathbf{L}}^{(d,m)}$, given by
\begin{equation}
\label{eq:Lhh}
\hat{L}^{(d,m)}_{is} = \sum_{\substack{i_2,\dots,i_{m-1}\\j_1,\dots,j_q}}A^{(d,m)}_{(i,s,i_2,\dots,i_{m-1})(j_1,\dots,j_q)} \quad \forall i\neq s \text{ and } \hat{L}^{(d,m)}_{ii} = -\sum_{s\neq i}\hat{L}^{(d,m)}_{is} \, ,
\end{equation}
and
\begin{equation}
\label{eq:Lht}
\check{L}^{(d,m)}_{is} = \sum_{\substack{i_1,\dots,i_{m-1}\\j_2,\dots,j_q}}A^{(d,m)}_{(i,i_1,\dots,i_{m-1})(s,j_2,\dots,j_q)} \quad \forall i\neq s \text{ and } \check{L}^{(d,m)}_{ii} = -\sum_{s\neq i}\check{L}^{(d,m)}_{is} \, .
\end{equation}
The former one is thus defined by considering indexes, $i$ and $s$, lying both in the head of the hyperedge, while for the latter, one index, $i$, is in the head and the second, $s$, in the tail. Because of the symmetry of the adjacency tensor with respect to a permutation of indexes in the head, the matrix $\hat{\;\mathbf{L}}^{(d,m)}$ is symmetric by construction. On the other hand, the matrix $\check{\;\mathbf{L}}^{(d,m)}$ is generally asymmetric and it carries the relevant information about the directionality of the hypergraph.

Let us observe that the latter matrices can be related to the higher-order Laplace matrix defined in~\cite{gallo2022synchronization}. For instance, let us define the generalized in-degree of node $i$ as the number of $m$-directed $d$-hyperedges containing $i$ in the head, namely
\begin{equation}
\label{eq:kin}
   k^{(d,m)}_{i} = \frac{1}{q!(m-1)!}\sum_{\substack{i_1,\dots,i_{m-1}\\j_1,\dots,j_q}}A^{(d,m)}_{(ii_1\dots i_{m-1})(j_1\dots j_q)} \, ,
\end{equation}
where the term $(m-1)!$ takes into account the permutation of the remaining $(m-1)$ nodes in the head, whereas $q!$ accounts for the permutation of tail nodes (remember $m+q=d+1$). Given a second node, $s$, we can then count the number of $m$-directed $d$-hyperedges containing both $i$ and $s$. While $i$ should always belong to the head, $s$ can be in the head or in the tail, we thus have
\begin{equation}
\label{eq:headkis}
   \hat{k}^{(d,m)}_{i,s} = \frac{1}{q!(m-2)!}\sum_{\substack{i_2,\dots,i_{m-1}\\j_1,\dots,j_q}}A^{(d,m)}_{(i,s,i_2,\dots,i_{m-1})(j_1,\dots,j_q)} \, ,
\end{equation}
in the case $s$ is an head node, and
\begin{equation}
\label{eq:tailkis}
   \check{k}^{(d,m)}_{i,s} = \frac{1}{(q-1)!(m-1)!}\sum_{\substack{i_1,\dots,i_{m-1}\\j_2,\dots,j_q}}A^{(d,m)}_{(i,i_1,\dots,i_{m-1})(s,j_2,\dots,j_q)} \, ,
\end{equation}
if $s$ belongs to the tail. Let us observe that we have
\begin{equation*}
\sum_s \hat{k}^{(d,m)}_{i,s}=(m-1) k^{(d,m)}_{i} \text{ and } \sum_s \check{k}^{(d,m)}_{i,s} = q k^{(d,m)}_{i}\,.
\end{equation*}
We can thus rewrite $\hat{\;\mathbf{L}}^{(d,m)}$ and $\check{\;\mathbf{L}}^{(d,m)}$ as follows
\begin{equation}
\label{eq:Lhh2}
\hat{L}^{(d,m)}_{is} = 
\begin{cases}
 q!(m-2)!\hat{k}^{(d,m)}_{i,s} & \text{ if $i\neq s$}\\
 - q!(m-1)!k^{(d,m)}_{i} & \text{ if $i= s$}\\
 0 & \text{ if $i$ or $s$ does not belong to the head}\, ,
\end{cases}
\end{equation}
and
\begin{equation}
\label{eq:Lht2}
\check{L}^{(d,m)}_{is} = 
\begin{cases}
  (q-1)!(m-1)!\check{k}^{(d,m)}_{i,s} & \text{ if $i\neq s$}\\
 - q!(m-1)!k^{(d,m)}_{i} & \text{ if $i= s$}\\
 0 & \text{ if $i$ does not belong to the head or if $s$ does not belong to the tail}\, .
\end{cases}
\end{equation}

In Fig.~\ref{fig:1directed3directed}, we show the Laplace matrices $\hat{\;\mathbf{L}}^{(d,m)}$ and $\check{\;\mathbf{L}}^{(d,m)}$ for some examples of $m$-directed $d$-hyperedges. One can appreciate that the matrices $\hat{\;\mathbf{L}}^{(d,1)}$ always vanish, $\hat{\;\mathbf{L}}^{(d,m)}$ are symmetric matrices while $\check{\;\mathbf{L}}^{(d,m)}$ are asymmetric ones.

In conclusion, by using the Laplace matrices $\hat{\;\mathbf{L}}^{(d,m)}$ and  $\check{\;\mathbf{L}}^{(d,m)}$, Eq.~\eqref{eq:manybodymDlin} can be rewritten as
\begin{equation}
\label{eq:manybodymDlinLap}
\frac{d\delta\vec{x}_i}{dt} = \mathbf{J}^{(0)}\delta\vec{x}_i+\sum_{d=1}^D \sigma_d \sum_{m=1}^d\sum_{s}\left[\hat{L}^{(d,m)}_{is}\left(\mathbf{J}_1 \vec{h}^{(d,m)}+\dots+ \mathbf{J}_{m-1} \vec{h}^{(d,m)}\right)\delta \vec{x}_{s}+\check{L}^{(d,m)}_{is}\left(\mathbf{J}_{m} \vec{h}^{(d,m)}+\dots+ \mathbf{J}_{d} \vec{h}^{(d,m)}\right)\delta \vec{x}_{s}\right]\, .
\end{equation}
By introducing the vector $\delta\vec{x}=\left(\delta\vec{x}_1^\top,\dots,\delta\vec{x}_N^\top\right)^\top$ describing the whole system state, the above equation can be rewritten in a more compact form:
\begin{equation}
\label{eq:manybodymDlinLapCompact}
\frac{d\delta\vec{x}}{dt} = \left[\mathbb{I}_N\otimes \mathbf{J}^{(0)}+\sum_{d=1}^D \sigma_d \sum_{m=1}^d \left(\hat{\;\mathbf{L}}^{(d,m)}\otimes \hat{\mathbf{J}}^{(d,m)} +\check{\;\mathbf{L}}^{(d,m)}\otimes \check{\mathbf{J}}^{(d,m)}\right)\right] \delta \vec{x}\, ,
\end{equation}
where $\mathbb{I}_N$ is the $N\times N$ identity matrix, $\otimes$ denotes the Kronecker product and the matrices 
\begin{equation}
    \label{eq:hatJcheckJ}
    \hat{\mathbf{J}}^{(d,m)}=\mathbf{J}_1 \vec{h}^{(d,m)}+\dots+ \mathbf{J}_{m-1} \vec{h}^{(d,m)} \quad\text{and}\quad\check{\mathbf{J}}^{(d,m)}=\mathbf{J}_{m} \vec{h}^{(d,m)}+\dots+ \mathbf{J}_{d} \vec{h}^{(d,m)}\, ,
\end{equation}
 are defined componentwise for all $i,j=1,\dots, n$ by
\begin{equation}
\label{eq:detJHH}
\hat{{J}}_{ij}^{(d,m)} = \frac{\partial h_i^{(d,m)}}{\partial x_{1,j}}+\dots+ \frac{\partial h_i^{(d,m)}}{\partial x_{m-1,j}} \text{ and }\check{{J}}_{ij}^{(d,m)}=\frac{\partial h_i^{(d,m)}}{\partial x_{m,j}}+\dots+ \frac{\partial h_i^{(d,m)}}{\partial x_{d,j}}\, ,
\end{equation}
where we emphasized the $n$ components of the vector function $\vec{h}^{(d,m)}=(h_1^{(d,m)},\dots,h_n^{(d,m)})^\top$ and of the vectors $\vec{x}_\ell = (x_{\ell,1},\dots, x_{\ell,n})^\top$, for $\ell=1,\dots,N$. We recall that the derivatives have been evaluated on the stationary solution $\vec{x}^*$.

To proceed within the standard framework allowing to obtain the dispersion relation to determine the onset of Turing patterns (or the Master Stability Function in the case of synchronization~\cite{PC1998,fujisaka1983stability}), one would like to determine a suitable basis to project Eq.~\eqref{eq:manybodymDlinLapCompact}, to reduce thus the dimension of the problem and hence making possible to achieve some analytical results. Here we are facing with some issues, often found in other works, namely there is not such a ``simple'' basis, because in general the matrices $\hat{\;\mathbf{L}}^{(d,m)}$ and $\check{\;\mathbf{L}}^{(d,m)}$ cannot be simultaneously diagonalized. Observe moreover that the latter is asymmetric and thus possibly not diagonalizable.

To overcome this issue scholars have considered two main cases: to introduce some assumptions on the hypergraph topology, e.g., deal with {\em regular topologies}~\cite{muolo2023turinghh,SayeedAnwar_2024} for which the Laplace matrices differ by a multiplicative constant, or the {\em natural coupling} assumption on the coupling functions resulting into a single Jacobian matrix that can thus be factorized. Let us observe that in the present case we will be dealing with both symmetric and asymmetric Laplace matrices and thus we can not use (in general) the regular topology assumption. For this reason we are considering in the following a slight generalization of the natural coupling assumption.

\subsection{Generalized natural coupling}
\label{ssec:gennatcoup}

The natural coupling assumption states that the coupling functions coincide for all order, $1\leq d \leq D$, once evaluated on the synchronous manifold, in formula (discarding the directionality $m$)
\begin{equation*}
 \vec{h}^{(d)}(\vec{x},\dots,\vec{x})=\dots =   \vec{h}^{(2)}(\vec{x},\vec{x})=\vec{h}^{(1)}(\vec{x}) \quad\forall \vec{x}\in\mathbb{R}^n\, ,
\end{equation*}
that leads to analogous equalities among the Jacobian matrices, namely
\begin{equation*}
\mathbf{J} \vec{h}^{(d)}=\dots =  \mathbf{J} \vec{h}^{(2)}=\mathbf{J} \vec{h}^{(1)}\, ,
\end{equation*}
where $\mathbf{J} \vec{h}^{(d)}=\mathbf{J}_1 \vec{h}^{(d)}+\dots+\mathbf{J}_d \vec{h}^{(d)}$. Let us observe that a similar conclusion would not be useful in our framework because the directionality induces a split of the derivatives determining two matrices $\hat{\mathbf{J}} \vec{h}^{(d,m)}$ and $\check{\mathbf{J}} \vec{h}^{(d,m)}$.

Our goal is thus to determine a condition on $\vec{h}^{(d,m)}$ to obtain $\hat{\mathbf{J}}^{(d,m)}=\alpha^{(d,m)}\check{\mathbf{J}}^{(d,m)}$ for some positive constant $\alpha^{(d,m)}$, in such a way Eq.~\eqref{eq:manybodymDlinLapCompact} rewrites
\begin{equation}
\label{eq:manybodymDlinLapCompact2}
\frac{d\delta\vec{x}}{dt} = \left[\mathbb{I}_N\otimes \mathbf{J}^{(0)}+\sum_{d=1}^D \sigma_d \sum_{m=1}^d \left(\alpha^{(d,m)}\hat{\;\mathbf{L}}^{(d,m)}+\check{\;\mathbf{L}}^{(d,m)}\right)\otimes \check{\mathbf{J}}^{(d,m)}\right] \delta \vec{x}\, .
\end{equation}
We can thus define an {\em effective} asymmetric Laplace matrix 
\begin{equation}
\label{eq:Mmatrix}
\mathbf{M}^{(d,m)}:=\alpha^{(d,m)}\hat{\;\mathbf{L}}^{(d,m)}+\check{\;\mathbf{L}}^{(d,m)}\, ,
\end{equation}
that rules the short time evolution of the perturbation $\delta\vec{x}$
\begin{equation}
\label{eq:manybodymDlinLapCompact3}
\frac{d\delta\vec{x}}{dt} = \left[\mathbb{I}_N\otimes \mathbf{J}^{(0)}+\sum_{d=1}^D \sigma_d \sum_{m=1}^d \mathbf{M}^{(d,m)}\otimes \check{\mathbf{J}}^{(d,m)}\right] \delta \vec{x}\, .
\end{equation}

For a sake of definitiveness we decided to present hereby the required condition in a simplified framework, leaving the full description to Appendix~\ref{sec:gencaseHcond}. Let us thus assume the functions $\vec{h}^{(d,m)}$ to depend only on one component of the vectors $\vec{x}_\ell\in\mathbb{R}^n$. Observe that this is the setting used in~\cite{HCLP2009} where the coupling function encodes the fact that the $j$--th component of one oscillator is linearly coupled to the $k$--th component of another oscillator, i.e., $h_k(\vec{x})=\delta_{ki}x_j$, or the nonlinear setting used in~\cite{gambuzza2021synchronization,gallo2022synchronization,muolo2023turinghh} where the coupling is given by, e.g., $\vec{h}(\vec{x})=(x_j^2x_k,0,0)$.

Let us thus assume $\vec{h}^{(d,m)}(\vec{x}_{\ell_1},\dots,\vec{x}_{\ell_d})$ to depend only on the $p$--th components of the vectors $\vec{x}_\ell$, namely $\vec{h}^{(d,m)}=\vec{h}^{(d,m)}(x_{{\ell_1},p},\dots,{x}_{{\ell_d},p})$. This implies that for all $\ell=1,\dots,d$
\begin{equation*}
 (\mathbf{J}_\ell \vec{h}^{(d,m)})_{ij}=\frac{\partial h_i^{(d,m)}}{\partial x_{\ell,j}}=\frac{\partial h_i^{(d,m)}}{\partial x_{\ell,j}}\delta_{jp}\, .
\end{equation*}
Moreover, we hypothesize the coupling function to reduce to a single monomial, namely ${h}_i^{(d,m)}= C^{(i)}_d x_{{\ell_1},p}^{a_1^{(i)}}\dots x_{{\ell_d},p}^{a^{(i)}_d}$ for some constant $C^{(i)}_d$, positive real numbers $a_1^{(i)},\dots , a^{(i)}_d$, and that there exists a constant $\alpha^{(d,m)}>0$ such that
\begin{equation}
\label{eq:conditionsimple}
 \alpha^{(d,m)} = \frac{a_1^{(i)}+\dots +a^{(i)}_{m-1}}{a_m^{(i)}+\dots + a^{(i)}_d}\, .
\end{equation}
Then, we claim that
\begin{equation}
\label{eq:condJJ}
\hat{\mathbf{J}}^{(d,m)} =\alpha^{(d,m)}\check{\mathbf{J}}^{(d,m)}  \, .
\end{equation}

To prove the above relation, let us determine the Jacobian matrices; for all $i,j=1,\dots,N$ we have
\begin{equation*}
 \hat{{J}}_{ij}^{(d,m)}=\sum_{\ell=1}^{m-1} (\mathbf{J}_\ell \vec{h}^{(d,m)})_{ij}=  \sum_{\ell=1}^{m-1}\frac{\partial h_i^{(d,m)}}{\partial x_{\ell,j}} \delta_{jp}\text{ and } \check{{J}}_{ij}^{(d,m)}=\sum_{\ell=m}^{d} (\mathbf{J}_\ell \vec{h}^{(d,m)})_{ij}=  \sum_{\ell=m}^{d}\frac{\partial h_i^{(d,m)}}{\partial x_{\ell,j}}\delta_{jp} \, .
\end{equation*}
We can hence explicitly compute the derivatives and by evaluating them on the equilibrium $\vec{x}_\ell=\vec{x}^*=(x_1^*,\dots,x_p^*,\dots,x_n^*)^\top$ for all $\ell=1,\dots,d$, we conclude
\begin{equation*}
 \hat{{J}}_{ij}^{(d,m)}=C^{(i)}_d(a_1^{(i)}+\dots +a^{(i)}_{m-1})(x_p^*)^{|a_\ell^{(i)}|-1} \text{ and } \check{{J}}_{ij}^{(d,m)}=C^{(i)}_d(a_m^{(i)}+\dots +a^{(i)}_{d})(x_p^*)^{|a_\ell^{(i)}|-1} \, ,
\end{equation*}
where we defined  $|a_\ell^{(i)}|=a_1^{(i)}+\dots +a^{(i)}_{d}$. Finally, by invoking~\eqref{eq:conditionsimple}, we have
\begin{equation*}
 \hat{{J}}_{ij}^{(d,m)}=C^{(i)}_d(a_1^{(i)}+\dots +a^{(i)}_{m-1})(x_p^*)^{|a_\ell^{(i)}|-1} = \alpha^{(d,m)} C^{(i)}_d(a_m^{(i)}+\dots +a^{(i)}_{d})(x_p^*)^{|a_\ell^{(i)}|-1}=\alpha^{(d,m)} \check{{J}}_{ij}^{(d,m)} \, .
\end{equation*}
Let us stress that, as previously observed, all the above conditions are trivially satisfied for $m=1$ because $\hat{\mathbf{J}}^{(d,1)}=0$.

To conclude this part, let us present few examples of functions satisfying the above condition.
\begin{remark}[Some examples]
\label{rem:example}
Functions for which the above condition holds true in the case $n=2$, i.e., $\vec{x}=(u,v)$, $p=1$ and $d=3$ are given by $\vec{h}^{(3,m)}(\vec{x}_{\ell_1},\vec{x}_{\ell_2},\vec{x}_{\ell_3})=({h}_1^{(3,m)}(u_{\ell_1},u_{\ell_2},u_{\ell_3}),0)^\top$ where~\footnote{We hereby assume for a sake of simplicity that $C^{(i)}_d=1$ and also $\alpha^{(d,m)}=1$. We also consider the function to a have a single non-zero component. None of those assumptions are necessary and can be relaxed without any loss of generality.}
\begin{equation}
 \label{eq:exampleh}
{h}_1^{(3,2)}(u_{\ell_1},u_{\ell_2},u_{\ell_3})=u_{\ell_1}^2u_{\ell_2}u_{\ell_3}\text{ and } {h}_1^{(3,3)}(u_{\ell_1},u_{\ell_2},u_{\ell_3})=u_{\ell_1}u_{\ell_2}u_{\ell_3}^2\, .
\end{equation}
Indeed, in the first case, $m=2$, we have
\begin{equation*}
 \frac{\partial{h}_1^{(3,2)}}{\partial u_{\ell_1}}=2 (u^*)^3= \frac{\partial{h}_1^{(3,2)}}{\partial u_{\ell_2}}+\frac{\partial{h}_1^{(3,2)}}{\partial u_{\ell_3}}\, ,
\end{equation*}
where $\vec{x}^*=(u^*,v^*)$, and thus
\begin{equation*}
\hat{\mathbf{J}}^{(3,2)} = \left(
\begin{matrix}
 2(u^*)^3 & 0\\
 0 & 0
\end{matrix}
 \right)=\check{\mathbf{J}}^{(3,2)} \, .
\end{equation*}
In the second case, $m=3$, we obtain
\begin{equation*}
 \frac{\partial{h}_1^{(3,3)}}{\partial u_{\ell_1}}+ \frac{\partial{h}_1^{(3,3)}}{\partial u_{\ell_2}}=(u^*)^3+(u^*)^3= \frac{\partial{h}_1^{(3,3)}}{\partial u_{\ell_3}}\, ,
\end{equation*}
 and eventually
\begin{equation*}
\hat{\mathbf{J}}^{(3,3)} = \left(
\begin{matrix}
 2(u^*)^3 & 0\\
 0 & 0
\end{matrix}
 \right)=\check{\mathbf{J}}^{(3,3)} \, .
\end{equation*}
Observe that
\begin{equation*}
\hat{\mathbf{J}}^{(3,1)} = 0 \text{ and }\check{\mathbf{J}}^{(3,1)}= \left(
\begin{matrix}
 2(u^*)^3 & 0\\
 0 & 0
\end{matrix}
 \right)\, ,
\end{equation*}
by taking for instance ${h}_1^{(3,1)}(u_{\ell_1},u_{\ell_2},u_{\ell_3})=\frac{1}{2}u_{\ell_1}^2u_{\ell_2}u_{\ell_3}$ and ${h}_2^{(3,1)}=0$.

In the case $d=2$ and still $p=1$ and $n=2$, we can use $\vec{h}^{(2,m)}(\vec{x}_{\ell_1},\vec{x}_{\ell_2})=({h}_1^{(2,m)}(u_{\ell_1},u_{\ell_2}),0)^\top$ with
\begin{equation*}
 {h}_1^{(2,2)}(u_{\ell_1},u_{\ell_2})=u_{\ell_1}^2u_{\ell_2}^2\, ,
\end{equation*}
that verifies
\begin{equation*}
 \frac{\partial{h}_1^{(2,2)}}{\partial u_{\ell_1}}=2(u^*)^3= \frac{\partial{h}_1^{(2,2)}}{\partial u_{\ell_2}}\, ,
\end{equation*}
returning
\begin{equation*}
\hat{\mathbf{J}}^{(2,2)} = \left(
\begin{matrix}
 2(u^*)^3 & 0\\
 0 & 0
\end{matrix}
 \right)=\check{\mathbf{J}}^{(2,2)} \, .
\end{equation*}
Once again, we get
\begin{equation*}
\hat{\mathbf{J}}^{(2,1)} = 0 \text{ and }\check{\mathbf{J}}^{(2,1)}= \left(
\begin{matrix}
 2(u^*)^3 & 0\\
 0 & 0
\end{matrix}
 \right)\, ,
\end{equation*}
by taking for instance ${h}_1^{(2,1)}(u_{\ell_1},u_{\ell_2})=\frac{1}{2}u_{\ell_1}^2u_{\ell_2}^2$ and ${h}_2^{(2,1)}=0$.

To conclude we can define $h^{(1,1)}_1=\frac{u_{\ell_1}^4}{2}$ to obtain 
\begin{equation*}
\check{\mathbf{J}}^{(1,1)} =  \left(
\begin{smallmatrix}
 2(u^*)^3 & 0\\
 0 & 0
\end{smallmatrix}
 \right)=\check{\mathbf{J}}^{(2,2)}=\check{\mathbf{J}}^{(2,1)}\, .
\end{equation*}
\end{remark}

Building on these examples we can conclude by saying that we can find constant $C^{(i)}_d$ and $\alpha^{(d,m)}$ such that Eq.~\eqref{eq:condJJ} is satisfied and moreover 
\begin{equation}
\label{eq:econJ}
\check{\mathbf{J}}^{(d,m)}=\check{\mathbf{J}}^{(d,m-1)}=\dots=\check{\mathbf{J}}^{(d,2)}=\dots =\check{\mathbf{J}}^{(1,1)}\, .
\end{equation}

Condition~\eqref{eq:econJ} is the required one to factorize the Jacobian of the coupling term in Eq.~\eqref{eq:manybodymDlinLapCompact} and thus to proceed with the standard strategy of projecting on the eigenbasis of a suitable non-symmetric effective Laplace matrix, $\mathbf{M}^{(d,m)}$, see Eq.~\eqref{eq:Mmatrix}, and eventually rewrite~\eqref{eq:manybodymDlinLapCompact3} as follows:
\begin{equation}
\frac{d\delta\vec{x}}{dt} = \left[\mathbb{I}_N\otimes \mathbf{J}^{(0)}+\mathbf{M}\otimes \check{\mathbf{J}}^{(1,1)}\right] \delta \vec{x}\, ,
\end{equation}
where $\mathbf{M}:=\sum_{d=1}^D \sigma_d \sum_{m=1}^d \mathbf{M}^{(d,m)}$.

The above matrix results to be a sort of effective Laplace matrix taking into account the hypergraph structure, the size and the directionality of hyperedges, as well as the coupling strength, $\sigma_d$, and it is thus generally an asymmetric matrix. Let us assume it admits an eigenbasis~\footnote{Observe that this condition can be relaxed and working with eigenvectors and generalized eigenvectors as shown in~\cite{dorchain2023defective} in the case of Turing patterns on defective networks.}, $\vec{\phi}^{(s)}$, associated to eigenvalues $\Lambda^{(s)}\in\mathbb{C}$, $s=1,\dots,N$. By projecting $\delta\vec{x}$ onto the latter, namely $\delta\vec{x}=\sum_{s=1}^N\eta_s \vec{\phi}^{(s)}$, we eventually obtain
\begin{equation}
\label{eq:manybodyMSE}
\frac{d\eta_s}{dt} =\left[\mathbf{J}^{(0)}+\Lambda^{(s)} \check{\mathbf{J}}^{(1,1)}\right] \eta_s \quad \forall s=1,\dots,N\, .
\end{equation}

In conclusion we have been able, in the spirit of the Turing theory on networks~\cite{NM2010}, to map the highly dimensional linear system~\eqref{eq:manybodymDlinLap} into a $1$-parameter family of smaller linear systems, from which the instability feature of the homogeneous equilibrium $\vec{x}^*$ can be analytically studied. Before studying the emergence of Turing patterns in the general case of $m$-directed $d$-hypergraphs, let us present the peculiar case $m=1$.

\subsection{$1$-directed hypergraphs}
\label{ssec:1dh}

Let us now consider the case of $1$-directed $d$-hypergraphs, namely for any hyperedge the head will contain a single node, $m=1$, and the remaining $d$ ones will form the tail. This implies that the coupling functions $\vec{h}^{(d,1)}$ will depend only on variables associated to tail nodes, indeed from $m+q=d+1$, it follows that $q=d$.

A straightforward consequence of this assumption is that the term involving $\hat{\;\mathbf{L}}^{(d,m)}$ in Eq.~\eqref{eq:manybodymDlinLap}  disappears and thus Eq.~\eqref{eq:manybodymDlinLapCompact} rewrites as
\begin{equation}
\label{eq:manybodymDlinLapCompactm1}
\frac{d\delta\vec{x}}{dt} = \left[\mathbb{I}_N\otimes \mathbf{J}^{(0)}+\sum_{d=1}^D \sigma_d \check{\;\mathbf{L}}^{(d,1)}\otimes \check{\mathbf{J}}^{(d,1)}\right] \delta \vec{x}\, .
\end{equation}
This is the same equation used in~\cite{gallo2022synchronization} in the framework of synchronization and thus one can proceed by assuming the natural coupling assumption for $\vec{h}^{(d,1)}$ or the regular topology hypothesis on the $1$-directed $d$-hypergraph.

\section{Turing patterns in $m$-directed $d$-hypergraphs}
\label{sec:TP}

The goal of this section is to specialize the general theory presented so far to the framework of Turing patterns that is often described for two species reaction-diffusion systems. Moreover, to emphasize the relevance of the $m$-directionality, we will assume that only $(d+1)$-interactions are allowed, namely we are dealing with $m$-directed $d$-hypergraphs. Let us observe that in the literature can be also found with the name $d$-uniform hypergraph, in the symmetric case. Let us stress that the developed theory goes beyond this case and can handle more general structures as we will show in Section~\ref{sec:dirtoundir}.

Having thus set $n=2$, hence $\vec{x}_i=(u_i,v_i)$, we can write Eq.~\eqref{eq:manybodymD} as follows
\begin{equation}
\label{eq:manybodymDTP}
\begin{cases}
\displaystyle\frac{d{u}_i}{dt} &= f_1^{(0)}\left({u}_i,{v}_i\right)+ \sigma_d  \displaystyle \sum_{
\substack{i_1,\dots,i_{m-1} \\ j_1,\dots,j_q}}\left[h_1^{(d,m)}\left({u}_{i_1},\dots,{u}_{i_{m-1}}{u}_{j_1},\dots,{u}_{j_q}\right)-h_1^{(d,m)}\left({u}_{i},\dots,{u}_{i}\right)\right] A^{(d,m)}_{(ii_1\dots i_{m-1})(j_1\dots j_q)}\\
\displaystyle\frac{d{v}_i}{dt} &= f_2^{(0)}\left({u}_i,{v}_i\right)+ \sigma_d  \displaystyle  \displaystyle \sum_{
\substack{i_1,\dots,i_{m-1} \\ j_1,\dots,j_q}} \left[h_2^{(d,m)}\left({v}_{i_1},\dots,{v}_{i_{m-1}}{v}_{j_1},\dots,{v}_{j_q}\right)-h_2^{(d,m)}\left({v}_{i},\dots,{v}_{i}\right)\right] A^{(d,m)}_{(ii_1\dots i_{m-1})(j_1\dots j_q)}\, ,
\end{cases}
\end{equation}
where we have defined $(f_1^{(0)},f_2^{(0)})=\vec{g}^{(0)}$, we have assumed $\vec{g}^{(d,m)}$ to be diffusive-like and $(h_1^{(d,m)},h_2^{(d,m)})=\vec{h}^{(d,m)}$. Moreover we make the hypothesis of absence of cross-coupling terms~\cite{muolo2023turinghh}, analogous to cross-diffusion terms, namely $h_1^{(d,m)}$ will depend only on the variable $u$ and $h_2^{(d,m)}$ on $v$.

By assuming the existence of a stable homogeneous equilibrium, $(u_i,v_i)=(u^*,v^*)$ for all $i=1,\dots,N$, namely 
\begin{equation*}
\mathbf{J}^{(0)}(u^*,v^*)=\left(
\begin{smallmatrix}
 \displaystyle \frac{\partial f_1^{(0)}}{\partial u} &  \displaystyle\frac{\partial f_1^{(0)}}{\partial v}\\
  \displaystyle\frac{\partial f_2^{(0)}}{\partial u} &  \displaystyle\frac{\partial f_2^{(0)}}{\partial v}
\end{smallmatrix}
\right)\, ,
\end{equation*}
is a stable matrix, one can perform the same strategy as in the general case, by defining $\delta u_i=u_i-u^*$, $\delta v_i=v_i-v^*$ and then by linearizing the nonlinear system~\eqref{eq:manybodymDTP}. One eventually obtains the analogue of Eq.~\eqref{eq:manybodymDlinLapCompact} (remember only $(d+1)$-body interactions are considered)
\begin{equation}
\label{eq:manybodymDlinLapCompactuv}
\frac{d\delta\vec{x}}{dt} = \left[\mathbb{I}_N\otimes \mathbf{J}^{(0)}+ \sigma_d  \left(\hat{\;\mathbf{L}}^{(d,m)}\otimes \hat{\mathbf{J}}^{(d,m)} +\check{\;\mathbf{L}}^{(d,m)}\otimes \check{\mathbf{J}}^{(d,m)}\right)\right] \delta \vec{x}\, ,
\end{equation}
where we have introduced the vector $\delta\vec{x}=\left(\delta u_1,\delta v_1, \dots , \delta u_N,\delta v_N\right)^\top$, the matrices $\hat{\;\mathbf{L}}^{(d,m)}$ and $\hat{\mathbf{J}}^{(d,m)}$ have been defined in Eqs.~\eqref{eq:Lhh2} and~\eqref{eq:Lht2} and the matrices $\hat{\mathbf{J}}^{(d,m)}$ and $\check{\mathbf{J}}^{(d,m)}$ defined in Eq.~\eqref{eq:detJHH} now read
\begin{equation}
\label{eq:detJHHTP}
\hat{\mathbf{J}}^{(d,m)} =
\left(\begin{smallmatrix}
 \frac{\partial h_1^{(d,m)}}{\partial u_{\ell_1}}+\dots+\frac{\partial h_1^{(d,m)}}{\partial u_{\ell_{m-1}}} &  0\\
  0 &  \frac{\partial h_2^{(d,m)}}{\partial v_{\ell_1}}+\dots+\frac{\partial h_2^{(d,m)}}{\partial v_{\ell_{m-1}}}
\end{smallmatrix}\right)\text{ and }\notag
\check{\mathbf{J}}^{(d,m)}=
\left(\begin{smallmatrix}
\frac{\partial h_1^{(d,m)}}{\partial u_{\ell_m}}+\dots+\frac{\partial h_1^{(d,m)}}{\partial u_{\ell_d}}& 0\\
 0 & \frac{\partial h_2^{(d,m)}}{\partial v_{\ell_m}}+\dots+\frac{\partial h_2^{(d,m)}}{\partial v_{\ell_d}}
\end{smallmatrix}\right)\, ,
\end{equation}
because of the assumption of absence of cross-diffusion to get $\frac{\partial h_1^{(d,m)}}{\partial v_i}=\frac{\partial h_2^{(d,m)}}{\partial u_i}=0$ for all $i=1,\dots,N$.

Under the hypothesis of general coupling presented in Section~\ref{ssec:gennatcoup} that we hereby assume to hold true, one can introduce the effective Laplace matrix $\mathbf{M}$. By assuming it admits an eigenbasis~\footnote{Let us observe that this assumption can be relaxed, we can deal with the generalized eigenvectors~\cite{dorchain2023defective} and get a similar result.}, one can project $\delta\vec{x}$ on such basis and obtain
\begin{equation*}
\frac{d\eta_s}{dt} =\left[\mathbf{J}^{(0)}+\sigma_d \Lambda^{(s)} \check{\mathbf{J}}^{(d,m)}\right] \eta_s:= \mathbf{J}_s \eta_s  \quad \forall s=1,\dots,N\, ,
\end{equation*}
where $\eta_s$ is the projection of $(\delta u_s , \delta v_s)^\top$ on the eigenbasis and $\Lambda^{(s)}$, $s=1,\dots,N$ the eigenvalues. The stability of the null solution of the previous equation is obtained by studying the characteristic equation $\det(\textbf{J}_s - \lambda_s \mathbb{I}_2)=0$, i.e.,
\begin{equation*}
    \lambda_s^2-\lambda_s \mathrm{tr}(\textbf{J}_s) + \det(\textbf{J}_s)=0 \,,
\end{equation*}
whose roots are given by
\begin{equation}
    \label{eq:RelDisp}
    \lambda_{s_{1,2}}=\frac{1}{2} \left( \mathrm{tr}(\textbf{J}_s) \pm \sqrt{\mathrm{tr}(\textbf{J}_s)^2 -4 \det(\textbf{J}_s)} \right) \, .
\end{equation}
The root with largest real part, seen as a function of the eigenvalues $\Lambda^{(s)}$, is the {\em dispersion relation}, i.e., $\lambda_s\equiv \lambda (\Lambda^{(s)})=\max \Re \lambda_{s_{1,2}}$. The system exhibits a Turing instability if there exists $s$ such that $\lambda_{s}$ is positive.

Because the hypergraphs we are considering are directed, the aggregated Laplace matrix we are working with is asymmetric, hence its eigenvalues are complex numbers, $\Lambda^{(s)}=\Re(\Lambda^{(s)})+ \iota\Im(\Lambda^{(s)})$, where $\iota=\sqrt{-1}$. The positivity of $\lambda_s$ can thus be studied by adapting the method developed in~\cite{asllani2014theory} to determine (in)stability region in the complex plane, the latter being defined via the inequality (see Appendix~\ref{sec:reldispcmplx})
\begin{equation}
\left[\Im\Lambda^{(s)}\right]^2    S_2\left( \Re\Lambda^{(s)}\right)< -S_1\left(\Re\Lambda^{(s)}\right)\, ,
\label{eq:instregcmplx}
\end{equation}
where $S_1$ and $S_2$ are polynomials given by
\begin{equation*}
    S_1\left( x\right) =C_{14}x^4+C_{13}x^3+C_{12}x^2+C_{11}x+C_{10}\text{  and  }
    S_2\left( x\right) =C_{22}x^2+C_{21}x+C_{20}\, .
\end{equation*}
The coefficients $C_{1j}$, $j=0,\dots,4$, and $C_{2j}$, $j=0,\dots,2$, can be explicitly computed (see Appendix~\ref{sec:reldispcmplx}) and depend on the model parameters, i.e., both the reaction and the coupling, but not on the hypergraph. Hence Eq.~\eqref{eq:instregcmplx} allows to completely disentangle the role of the hypergraph support, captured by the eigenvalues $\Lambda^{(s)}$, and of the dynamical part, i.e., the polynomials $S_1$ and $S_2$.

To study the impact of the directionality, i.e., $m$, on the onset of Turing patterns we proceed  with a concrete model. We selected the Brusselator reaction system~\cite{PrigogineNicolis1967,PrigogineLefever1968} that has been largely studied in the framework of Turing patterns both on networks~\cite{NM2010,BullaraDeDeckerLefever2013,asllani2014theory} and on higher-order structures~\cite{CarlettiFanelliNicoletti2020,muolo2023turinghh}. The Brusselator is a nonlinear model defined by
\begin{equation}
\label{eq:brusselator}
\begin{cases}
 \displaystyle\frac{du}{dt} &=1-(b+1)u+c u^2v\\
 \displaystyle  \frac{dv}{dt} &=bu-cu^2v\, ,
\end{cases}
\end{equation}
where $b$ and $c$ act as tunable non-negative parameters.

For a sake of definitiveness and to emphasize the role of the directionality, we considered a peculiar $m$-directed $d$-hyperring, namely an hypergraph where $Q$ hyperedges of size $d+1$ are arranged in a circular way, where one of the $m$ nodes in the head is also among the tail nodes of the successive hyperedge considered, say, in clockwise way (see left panels of Fig.~\ref{fig:d2m12Motif} and Fig.~\ref{fig:d3m123Motif} for few examples). Let us observe that this is the natural extension to the $m$-directed case of the symmetric hyperring defined in~\cite{MNGCF2024}. In particular we fixed $d=2$, $Q=15$ hyperedges, hence $N=30$ nodes in total, and we let $m$ to assume the values $1$ and $2$, observe thus that $q$ varies accordingly. To improve the visibility, we show on the left columns smaller hyperrings composed by only $Q=5$ hyperedges. The coupling functions are similar to the ones presented in Remark~\ref{rem:example} namely
$\vec{h}^{(2,2)}=(D_u u_{\ell_1}^2u_{\ell_2}^2,D_v v_{\ell_1}^2v_{\ell_2}^2)^\top$ and $\vec{h}^{(2,1)}=(D_u u_{\ell_1}^2u_{\ell_2}^2/2,D_v v_{\ell_1}^2v_{\ell_2}^2/2)^\top$, where $D_u$ and $D_v$ are the diffusion coefficients of species $u$ and $v$. Remember that the Brusselator is a $2$-dimensional system, hence $n=2$. The involved Jacobian matrices are thus
\begin{equation*}
\hat{\mathbf{J}}^{(2,1)} =0\text{ and }\notag
\check{\mathbf{J}}^{(2,1)}=
\left(\begin{smallmatrix}
\frac{\partial h_1^{(2,1)}}{\partial u_{\ell_1}}+\frac{\partial h_1^{(2,1)}}{\partial u_{\ell_2}}& 0\\
 0 & \frac{\partial h_2^{(2,1)}}{\partial v_{\ell_1}}+\frac{\partial h_2^{(2,1)}}{\partial v_{\ell_2}}
\end{smallmatrix}\right)=
\left(\begin{matrix}
2D_u(u^*)^3& 0\\
 0 & 2D_v(v^*)^3
\end{matrix}\right)\, ,
\end{equation*}
and
\begin{equation*}
\hat{\mathbf{J}}^{(2,2)} =
\left(\begin{smallmatrix}
 \frac{\partial h_1^{(2,2)}}{\partial u_{\ell_1}} &  0\\
  0 &  \frac{\partial h_2^{(2,2)}}{\partial v_{\ell_1}}
\end{smallmatrix}\right)=
\left(\begin{matrix}
 2D_u(u^*)^3 &  0\\
  0 &  2D_v(v^*)^3
\end{matrix}\right)\text{ and }\notag
\check{\mathbf{J}}^{(2,2)}=
\left(\begin{smallmatrix}
\frac{\partial h_1^{(2,2)}}{\partial u_{\ell_2}}& 0\\
 0 & \frac{\partial h_2^{(2,2)}}{\partial v_{\ell_2}}
\end{smallmatrix}\right)=
\left(\begin{matrix}
 2D_u(u^*)^3 &  0\\
  0 &  2D_v(v^*)^3
\end{matrix}\right)\, .
\end{equation*}
In conclusion we have
\begin{equation*}
\hat{\mathbf{J}}^{(2,2)}= \check{\mathbf{J}}^{(2,2)}=\check{\mathbf{J}}^{(2,1)} \text{ and }\hat{\mathbf{J}}^{(2,1) }=0\, ,
\end{equation*}
and thus the matrices~\eqref{eq:Mmatrix} are given by
\begin{equation*}
\mathbf{M}^{(2,1)}=\check{\;\mathbf{L}}^{(2,1)}\text{ and } \mathbf{M}^{(2,2)}=\hat{\;\mathbf{L}}^{(2,2)}+\check{\;\mathbf{L}}^{(2,2)}\, ,
\end{equation*}
where $\hat{\;\mathbf{L}}^{(2,2)}$, $\check{\;\mathbf{L}}^{(2,1)}$ and $\check{\;\mathbf{L}}^{(2,1)}$ are the Laplace matrices of the underlying $m$-directed $2$-hyperring.

By exploiting the structure of $\mathbf{M}^{(2,1)}$ we can show that its spectrum, once we assume that $Q$ hyperedges are present, is given by
\begin{equation*}
    \Lambda^{(j)}=0\quad \forall j\in\{1,\dots,Q\}\text{ and }\Lambda^{(j+Q)}=-2+e^{2\pi i (j-1)/Q}\quad \forall j\in\{1,\dots,Q\}\, ,
\end{equation*}
namely $0$ has multiplicity $Q$ and the remaining eigenvalues recall the circular structure of the hyperring. Similarly, the spectrum of $\mathbf{M}^{(2,2)}$ is
\begin{equation*}
    \Lambda^{(j)}=-1+e^{2\pi i (j-1)/Q}\quad \forall j\in\{1,\dots,Q\}\text{ and }\Lambda^{(j+Q)}=-3\quad \forall j\in\{1,\dots,Q\}\, ,
\end{equation*}
in this case $0$ has multiplicity $1$, the eigenvalue $-3$ has multiplicity $Q$ and the remaining eigenvalues recall again the circular structure of the hyperring. We can thus conclude that once $m$ passes from $1$ to $2$ a ``large part'' of the spectrum moves toward the imaginary axis (see panels in the second column from the left of Fig.~\ref{fig:d2m12Motif}).

The results for the $m$-directed $2$-hyperring are reported in Fig.~\ref{fig:d2m12Motif}. Top row panels refer to the undirected case, middle row panels to $m=1$ while bottom row panels to $m=2$. Panels $b_1)$, $b_2)$ and $b_3)$ show the instability region in the complex plane where the inequality~\eqref{eq:instregcmplx} is satisfied (green region), black dots are the complex eigenvalues $\Lambda^{(s)}$ of the matrix $\mathbf{M}^{(d,m)}$; panels $c_1)$, $c_2)$ and $c_3)$ present the dispersion relation $\lambda_s$ as a function of $-\Re \Lambda^{(s)}$ (red dots); right panels the time evolution of $u_i(t)$. One can observe (top row panels) that for the undirected hypergraph the spectrum is real and thus it completely lies in the white region associated thus to a negative dispersion relation and no patterns can emerge. In the middle row panels, $m=1$, the spectrum is complex but it still lies in the white region, hence the same conclusions can be drawn, i.e., the homogeneous equilibrium remains stable and the patterns cannot develop. On the other hand, once $m=2$ (bottom row panels) the presence of several eigenvalues inside the instability region returns thus a positive dispersion relation (red dots with positive ordinate) and hence the emergence of patterns. We can conclude that directionality plays a role in patterns emergence: by increasing $m$ we can let the patterns to emerge, the reason being that the real part of a large number of eigenvalues increase and thus they enter the instability region that is located close to the imaginary axis.
\begin{figure}[h!]
    \centering
    \includegraphics[width=18cm ]{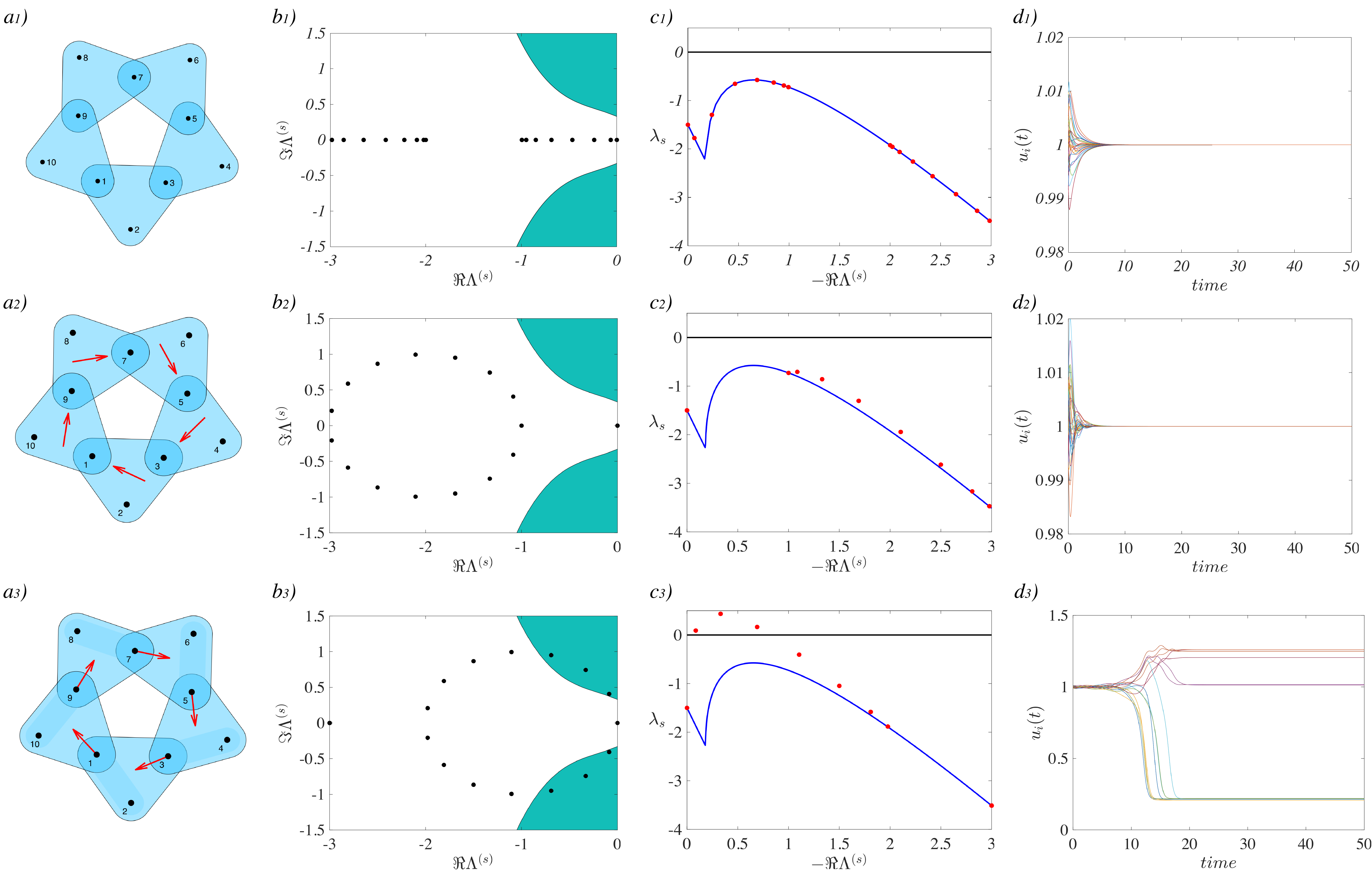}
    \caption{Undirected (top panels) and $m$-directed $2$-hyperring, $m=1$ (middle panels) and $m=2$ (bottom panels). On the left columns we show, for visualization purpose, a representative hyperring composed by $5$ hyperedges each one containing $3$ nodes, $m$ of which form the head and $q=3-m$ determine the tail, for a total of $10$ nodes, $m=1$ (panel $a_2$)) while $m=2$ (panel $a_3$). The results of remaining panels have been obtained by using a $2$-hyperring composed by $Q=15$ hyperedges and thus $N=30$ nodes. The hyperedge heads are emphasized in darker blue and the arrows help the reader to visualize the directionality. In panel $a_1)$ there is no directionality. Panels $b_1)$, $b_2)$ and $b_3)$ present the instability region in the complex plane (green area) and the complex spectrum of the effective Laplace matrix $\mathbf{M}^{(d,m)}$ (black dots). Panels $c_1)$, $c_2)$ and $c_3)$ report the dispersion relation $\lambda_s$ as a function of $-\Re\Lambda^{(s)}$. The time evolution of $u_i(t)$ is presented in panels $d_1)$, $d_2)$ an $d_3)$. The parameter of the Brusselator model are given by $b=5$, $c=7$ and the coupling functions are given by $h_1^{(2,1)}(u_1,u_2)=\frac{D_u}{2}u_1^2u_2^2$, $h_2^{(2,1)}(v_1,v_2)=\frac{D_v}{2}v_1^2v_2^2$, $h_1^{(2,2)}(u_1,u_2)=D_u u_1^2u_2^2$ and $h_2^{(2,2)}(v_1,v_2)=D_v v_1^2v_2^2$, with $D_u=1$, $D_v=9$, moreover $\sigma_2=1$. In the undirected case, we used ${h}_1^{(2)}(u_{1},u_{2})=\frac{D_u}{2}u_1^2u_2^2$, ${h}_2^{(2)}(v_{1},v_{2})=\frac{D_v}{2}v_1^2v_2^2$.}
    \label{fig:d2m12Motif}
\end{figure}

A similar result can be obtained by using $m$-directed $3$-hyperring formed by $Q=15$ hyperedges and thus $N=45$ nodes (see Fig.~\ref{fig:d3m123Motif}). In this case the coupling functions have been set as follows
\begin{equation*}
 \vec{h}^{(3,1)}=\frac{1}{2}\left(D_u u_{\ell_1}^2u_{\ell_2}u_{\ell_3},D_v v_{\ell_1}^2v_{\ell_2}v_{\ell_3}\right)^\top\, , \vec{h}^{(3,2)}=\left(D_u u_{\ell_1}^2u_{\ell_2}u_{\ell_3},D_v v_{\ell_1}^2v_{\ell_2}v_{\ell_3}\right)^\top\text{ and } \vec{h}^{(3,3)}=\left(D_u u_{\ell_1}u_{\ell_2}u^2_{\ell_3},D_v v_{\ell_1}v_{\ell_2}v^2_{\ell_3}\right)^\top\, ,
\end{equation*}
in such a way to obtain
\begin{equation*}
\hat{\mathbf{J}}^{(3,1)} = 0\,, \check{\mathbf{J}}^{(3,1)} = \check{\mathbf{J}}^{(3,2)}=\check{\mathbf{J}}^{(3,3)}=\hat{\mathbf{J}}^{(3,2)}=\hat{\mathbf{J}}^{(3,3)}=
\left(\begin{matrix}
 2D_u(u^*)^2 &  0\\
  0 &   2D_v(v^*)^2
\end{matrix}\right)\, .
\end{equation*}

By exploiting the structure of  $\mathbf{M}^{(3,1)}$ we can show that its spectrum is given by
\begin{equation*}
    \Lambda^{(j)}=0\quad \forall j\in\{1,\dots,2Q\}\text{ and }\Lambda^{(j+2Q)}=-6+2e^{2\pi i (j-1)/Q}\quad \forall j\in\{1,\dots,Q\}\, ;
\end{equation*}
the spectrum of $\mathbf{M}^{(3,2)}$ is
\begin{equation*}
    \Lambda^{(j)}=0\quad \forall j\in\{1,\dots,Q\}\, , \Lambda^{(j+Q)}=-6\quad \forall j\in\{1,\dots,Q\}\text{ and }\Lambda^{(j+2Q)}=-2+e^{2\pi i (j-1)/Q}\quad \forall j\in\{1,\dots,Q\}\, .
\end{equation*}
Finally the spectrum of $\mathbf{M}^{(3,3)}$ is
\begin{equation*}
    \Lambda^{(j)}=-2+2e^{2\pi i (j-1)/Q}\quad \forall j\in\{1,\dots,Q\}\text{ and }\Lambda^{(j+Q)}=-5\quad \forall j\in\{1,\dots,2Q\}\, .
\end{equation*}
We can thus conclude that also in this case, as we increase $m$ from $1$ to $3$ a ``large part'' of the spectrum moves toward the imaginary axis (see panels in the second column from the left of Fig.~\ref{fig:d3m123Motif}).

By varying $m$ into $\{1,2,3\}$ ($m=1$ second row panels from the top, $m=2$ third row panels from the top and $m=3$ bottom row panels) we can observe again that patterns emerge for large enough $m$, i.e., $m=3$, while they are not present for $m=1$ and $m=2$. In the undirected case (top row panels), patterns cannot develop because the real spectrum does not intersect the instability region in the complex plane.
\begin{figure}[h!]
    \centering
    \includegraphics[width=18cm ]{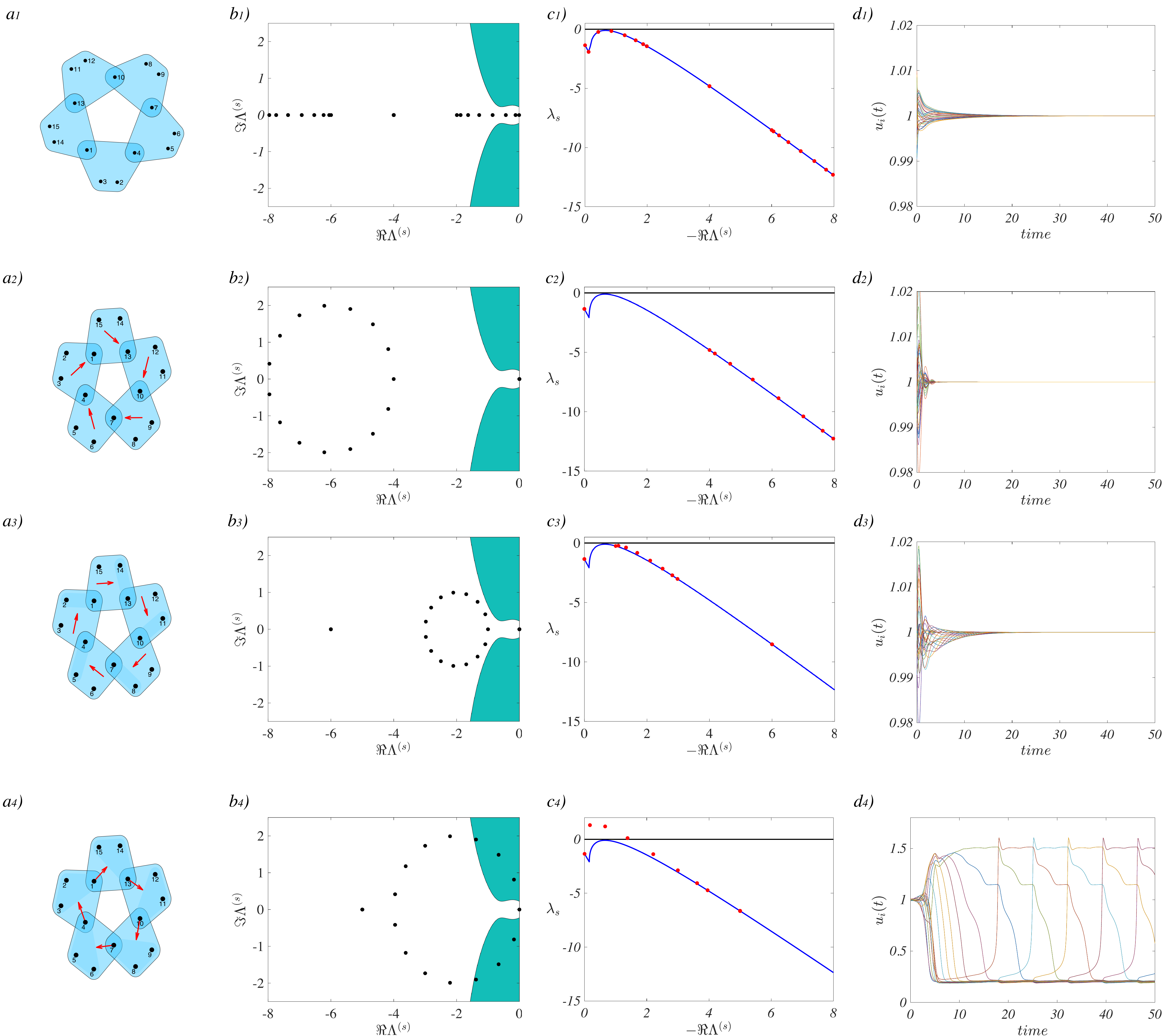}
    \caption{Undirected (top panels) and $m$-directed $3$-hyperring, $m=1$ (second panels from the top), $m=2$ (third panels from the top) and $m=3$ (bottom panels). On the left columns we show, for visualization purpose, a representative hyperring, composed by $5$ hyperedges each one containing $4$ nodes, $m$ of which form the head and $q=4-m$ determine the tail, for a total of $15$ nodes, $m=1$ (panel $a_2)$, $m=2$ (panel $a_3)$) and $m=3$ (panel $a_4)$). The results in the remaining panels have been obtained by using a $3$-hyperring composed by $Q=15$ hyperedges and thus $N=45$ nodes. The hyperedge heads are emphasized in darker blue and the arrows help the reader to determine the directionality. In panel $a_1)$ we show the undirected hyperring. Panels $b_1)$, $b_2)$, $b_3)$ and $b_4)$ present the instability region in the complex plane (green area) and the complex spectrum of the effective Laplace matrix $\mathbf{M}^{(d,m)}$ (black dots). Panels $c_1)$, $c_2)$, $c_3)$ and $c_4)$ report the dispersion relation $\lambda_s$ as a function of $-\Re\Lambda^{(s)}$. The time evolution of $u_i(t)$ is presented in panels $d_1)$, $d_2)$, $d_3)$ and $d_4)$. The parameter of the Brusselator model are given by $b=5.3$, $c=7$, $D_u=1$, $D_v=9$ and the coupling functions are given by $h_1^{(3,1)}(u_1,u_2,u_3)=\frac{D_u}{2}u_1^2u_2u_3$,  $h_2^{(3,1)}(v_1,v_2,v_3)=\frac{D_v}{2}v_1^2v_2v_3$, $h_1^{(3,2)}(u_1,u_2,u_3)=D_u u_1^2u_2u_3$, $h_2^{(3,2)}(v_1,v_2,v_3)=D_v v_1^2v_2v_3$,  $h_1^{(3,3)}(u_1,u_2,u_3)=D_u u_1u_2u_3^2$ and $h_2^{(3,3)}(v_1,v_2,v_3)=D_v v_1v_2v_3^2$ moreover $\sigma_2=1$. In the undirected case, we used ${h}_1^{(3)}(u_{1},u_{2},u_{3})=D_u u_{1}u_{2}u_{3}^{2}$, ${h}_2^{(3)}(v_{1},v_{2},v_{3})=D_v v_{1}v_{2}v_{3}^{2}$  as coupling functions.}
    \label{fig:d3m123Motif}
\end{figure}

Those preliminary results (see also Fig.~\ref{fig:d3m123Motifbis}) support thus the claim that patterns formation is enhanced by increasing the directionality, i.e., $m$. To strengthen this claim we will hereby consider the emergence of patterns in function of the model parameters $(b,c)$ (see Fig.~\ref{fig:directvariationbc} left panel) or the diffusion coefficients $(D_u,D_v)$ (see Fig.~\ref{fig:directvariationbc} right panel). The results reported in the latter figure show that $3$-directed $3$-hyperring will support Turing patterns for a much larger set of parameters $(b,c)$ (green-orange-yellow area) than the $2$-directed (orange-yellow area), that in turn will exhibit patterns for a choice of parameters $(b,c)$ larger than the $1$-directed case (yellow). Interestingly enough, the region in the $(b,c)$ plane, bounded by the black line, for which the undirected $3$-hyperring supports patterns is smaller than the region for which the $3$-directed $3$-hyperring exhibits patterns, i.e., the green-orange-yellow area. A similar conclusion can be drawn in the case of patterns emergence as a function of the diffusion coefficients $(D_u,D_v)$.
\begin{figure}[h!]
    \centering
    \includegraphics[width=8cm ]{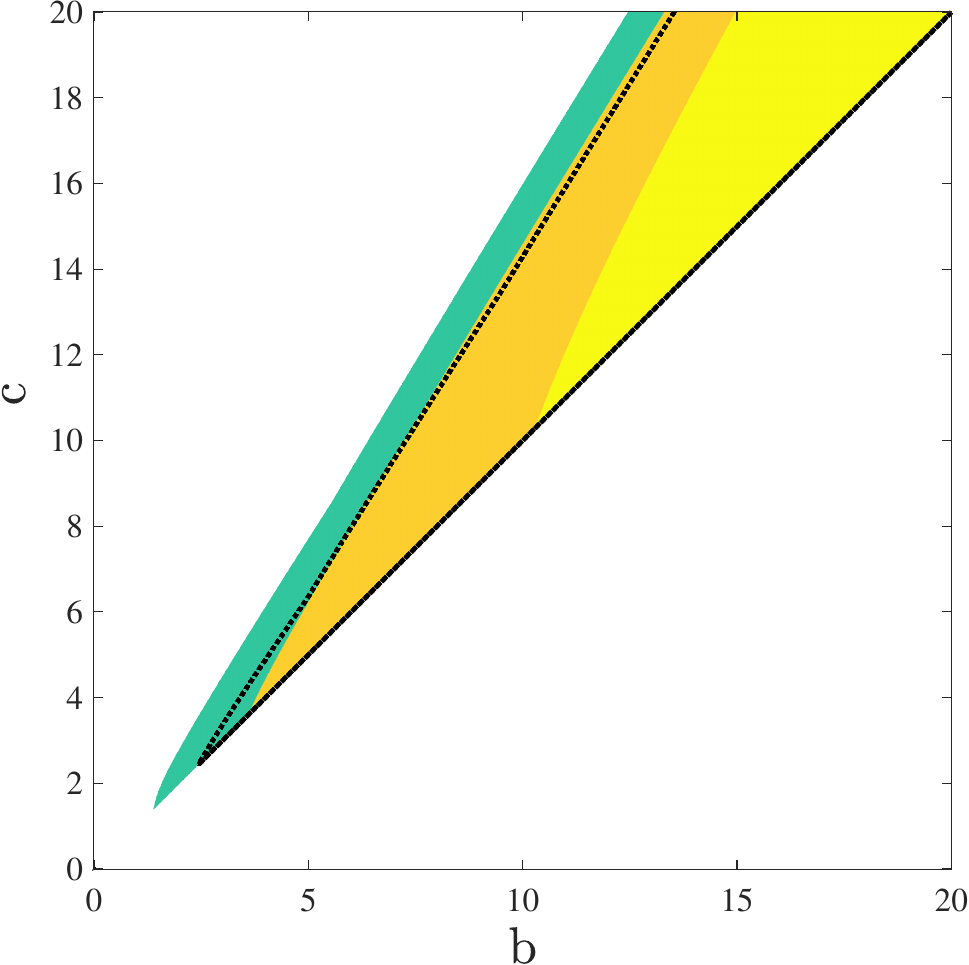}\quad
    \includegraphics[width=8cm ]{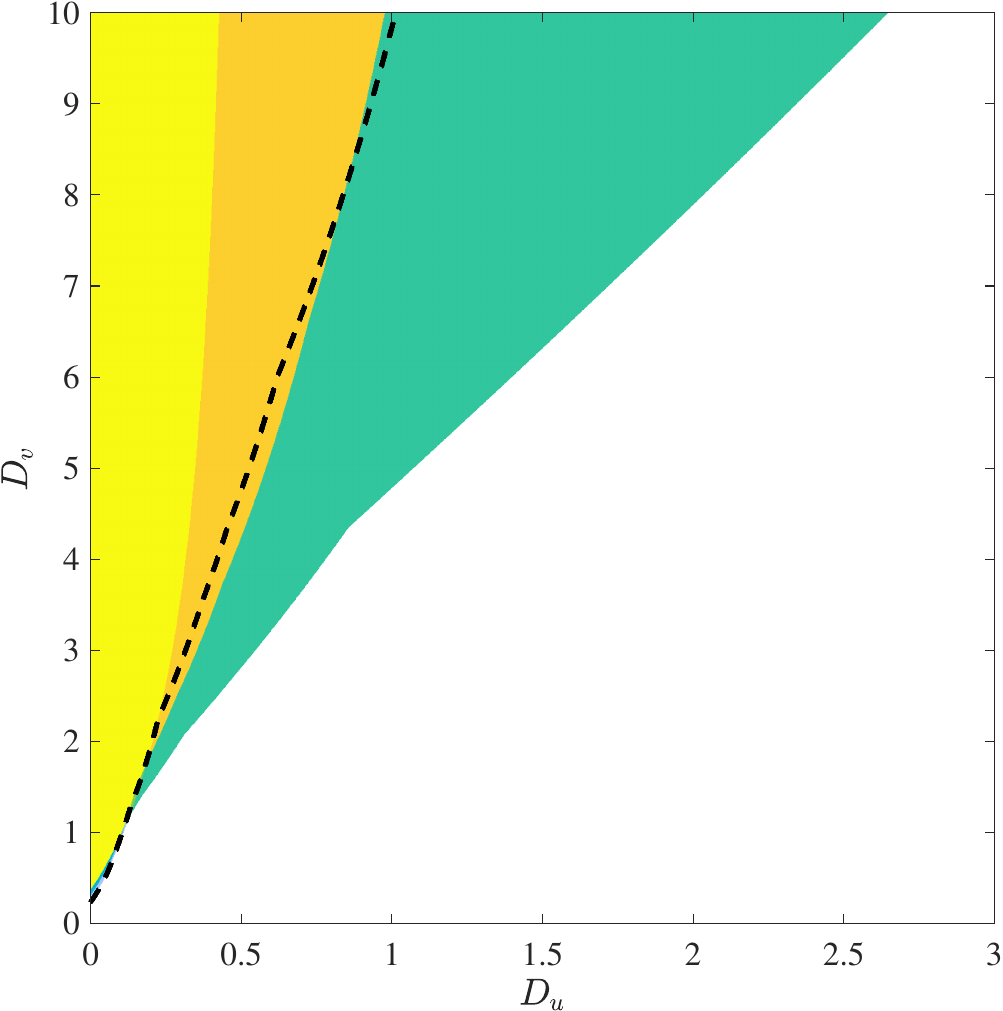}
    \caption{Directionality and emergence of patterns. We report the maximum of the dispersion relation as a function of the parameters $b$ and $c$ (left panel for $D_{u}=1$ and $D_{v}=9$) and the same quantity as a function of $D_u$ and $D_v$ (right panel for $b=5.3$ and $c=7$) by assuming the underlying hypergraph to be a $3$-hyperring composed by $Q=15$ hyperedges and thus $N=45$ nodes. The white region corresponds to a negative dispersion relation and thus to absence of patterns. The colored areas are associated to a positive maximum and thus to the presence of Turing patterns. More precisely, parameters $(b,c)$ or $(D_u,D_v)$ in the yellow region return patterns for $1$, $2$ and $3$-directed hyperring, the orange region denotes presence of patterns for $2$ and $3$-directed hyperring, while the green one only for $3$-directed hyperring. The region bounded by the black curve is associated to patterns for the undirected hyperring. The coupling functions are the same used in Fig.~\ref{fig:d3m123Motif}.} 
    \label{fig:directvariationbc}
\end{figure}

\section{Turing patterns from $m$-directed to undirected $d$-hypergraphs}
\label{sec:dirtoundir}

To continue the study on the impact of the directionality onto patterns emergence, we consider a model of hypergraph depending on a parameter allowing us to interpolate between a direct hypergraph and an undirected one. This idea is analogous to consider a couple of nodes in a directed network connected with two reciprocal links, one of weight $p$ and the other $1-p$, in fact by varying $p\in [0,1/2]$ one can evolve from a directed link to an undirected one. A similar strategy has been proposed in~\cite{gallo2022synchronization} in the case of $1$-directed hypergraphs, let us notice that the construction hereby presented generalizes the former along two directions, first by considering $m>1$ and second by taking into account the nature of the two involved Laplace matrices.

Let us denote by $p$ the parameter allowing to pass from a $m$-directed hypergraph, say $p=0$, to a symmetric one, say $p^*$. Because Turing instability is rooted on a linear stability analysis and to avoid confounding factors (mainly the generalized node degree), one would like the linearized systems for the directed hypergraph to reduce to the undirected case for $p\rightarrow p^*$. In the case $m=1$ this can be achieved by imposing the Laplace matrix of the $1$-directed hypergraph to coincide in this limit with the one of the undirected case~\cite{gallo2022synchronization}. If $m>1$, based on~\eqref{eq:Mmatrix}, we should require the matrix $\mathbf{M}^{(d,m)}$, that depends on the parameter $p$, to converge to the Laplace matrix of the undirected hypergraph, formally $\mathbf{M}^{(d,m)}(p)\rightarrow \mathbf{L}^{(d)}$ as $p\rightarrow p^*$. Let us observe that the matrix $\mathbf{M}^{(d,m)}(p)$ depends also on the coupling function via the coefficient $\alpha^{(d,m)}$.

For a sake of pedagogy let us consider a $m$-directed hypergraph formed by a single $d$-hyperedge and to fix notations, let us set $d=3$ and consider the case $m=3$ (see Fig.~\ref{fig:Fig4New}). Lets $q_i$, $i=1,\dots,4$, be four positive real numbers and consider the four basic $3$-directed $3$-hyperedges, then we can define the hyperedge obtained by juxtaposing the former hyperedges each one weighted by one of the $q_i$. A straightforward computation (see Eqs.~\eqref{eq:Lhh2} -~\eqref{eq:Lht2} and panel c) Fig.~\ref{fig:1directed3directed}) allows to obtain the matrices $\hat{\;\mathbf{L}}^{(3,3)}$ and $\check{\;\mathbf{L}}^{(3,3)}$ associated to the weighted hyperedge as follows
\begin{eqnarray*}
\hat{\;\mathbf{L}}^{(3,3)}& =& q_1\left(
\begin{matrix}
 0 & 0 & 0 & 0\\
 0 & -2 & 1 & 1\\
 0 & 1 & -2 & 1\\
 0 & 1 & 1 & -2
 \end{matrix}\right)+q_2\left(
\begin{matrix}
 -2 & 0 & 1 & 1\\
  0 & 0 & 0 & 0\\
  1 & 0 & -2 & 1\\
  1 & 0 & 1 & -2
\end{matrix}\right)+q_3\left(
\begin{matrix}
-2 & 1 & 0 & 1\\
1 & -2 & 0 & 1\\
0 & 0 & 0 & 0\\
1 & 1 & 0 & -2
\end{matrix}\right)+q_4\left(
\begin{matrix}
-2 & 1 & 1 & 0\\
1 & -2 & 1 & 0\\
1 & 1 & -2 & 0\\
0 & 0 & 0 & 0
\end{matrix}\right)\\
&=&\left(
\begin{matrix}
-2(q_2+q_3+q_4) & q_3+q_4 & q_2+q_4 & q_2+q_3\\
q_3+q_4 & -2(q_1+q_3+q_4) & q_1+q_4 & q_1+q_3\\
q_2+q_4 & q_1+q_4 & -2(q_1+q_2+q_4) & q_1+q_2\\
q_2+q_3 & q_1+q_3 & q_1+q_2 & -2(q_1+q_2+q_3)
\end{matrix}\right)\, ,
\end{eqnarray*}
and
\begin{eqnarray*}
\check{\;\mathbf{L}}^{(3,3)} &= &q_1\left(
\begin{matrix}
0 & 0 & 0 & 0\\
2 & -2 & 0 & 0\\
2 & 0 & -2 & 0\\
2 & 0 & 0 & -2
\end{matrix}\right)+q_2\left(
\begin{matrix}
-2 & 2 & 0 & 0\\
 0 & 0 & 0 & 0\\
 0 & 2 & -2 & 0\\
 0 & 2 & 0 & -2
\end{matrix}\right)+q_3\left(
\begin{matrix}
-2 & 0 & 2 & 0\\
0 & -2 & 2 & 0\\
0 & 0 & 0 & 0\\
0 & 0 & 2 & -2
\end{matrix}\right)+q_4\left(
\begin{matrix}
-2 & 0 & 0 & 2\\
0 & -2 & 0 & 2\\
0 & 0 & -2 & 2\\
0 & 0 & 0 & 0
\end{matrix}\right)\\
&=&\left(
\begin{matrix}
-2(q_2+q_3+q_4) & 2q_2 & 2q_3 & 2q_4\\
2q_1 & -2(q_1+q_3+q_4) & 2q_3 & 2q_4\\
2q_1 & 2q_2 & -2(q_1+q_2+q_4) & 2q_4\\
2q_1 & 2q_2 & 2q_3 & -2(q_1+q_4+q_3)
\end{matrix}\right)\, .
\end{eqnarray*}
\begin{figure}[h!]
    \centering
    \includegraphics[width=14cm ]{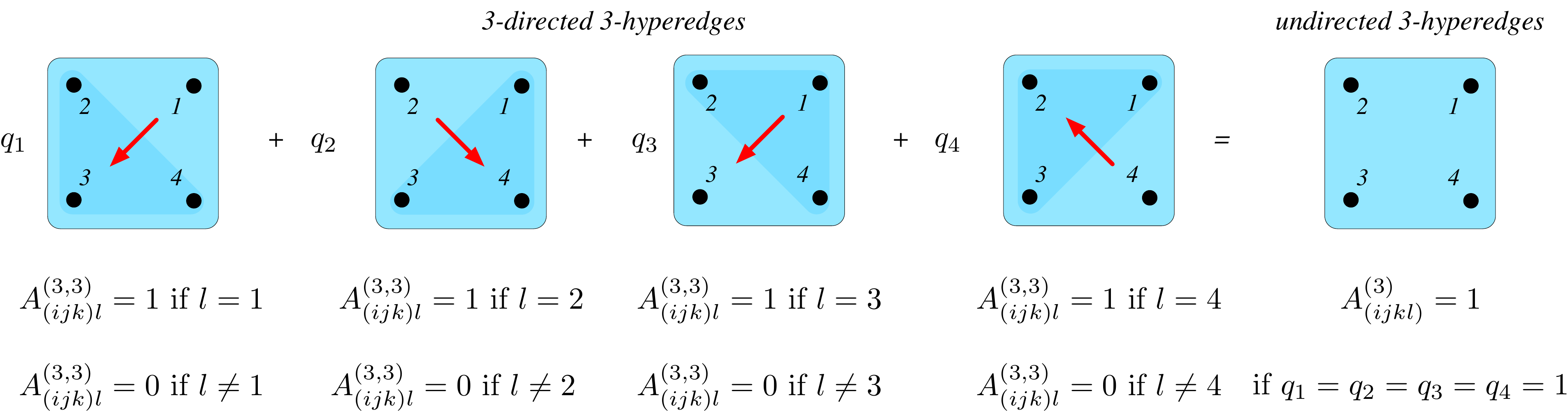}
    \caption{$3$-directed versus undirected $3$-hyperedges. We show how to obtain the undirected $3$-hyperedge by ``juxtaposition'' of four $3$-directed hyperedges weighted with positive coefficients $q_1=q_2=q_3=q_4=1/4$. For the remaining values of the latter, we obtain a directed $3$-hyperedge. The hyperedge heads are emphasized in darker blue and the arrows help the reader to determine the directionality.}
    \label{fig:Fig4New}
\end{figure}
    
By using again coupling functions such that $\hat{\mathbf{J}}^{(3,3)}=\check{\mathbf{J}}^{(3,3)}$, it results that $\mathbf{M}^{(3,3)}=\hat{\;\mathbf{L}}^{(3,3)}+\check{\;\mathbf{L}}^{(3,3)}$. The latter matrix will coincide with the Laplace matrix associated to the undirected $3$-hyperedge if $q_1=q_2=q_3=q_4=\frac{1}{4}$, the latter being $\mathbf{L}^{(3)}=\left(\begin{smallmatrix}
    -3 & 1 & 1 &1\\
    1 & -3 & 1 &1\\
    1 & 1 & -3 &1\\
    1 & 1 & 1 &-3
\end{smallmatrix}\right)$.

Let us observe that if on the other hand $\hat{\mathbf{J}}^{(d,m)}$ is proportional to $\check{\mathbf{J}}^{(d,m)}$, then this constraint should be taken into account to study the limit case. Let us conclude by remarking that given an undirected $d$-hyperedge there are $\binom{m}{d+1}$ ``elementary'' $m$-directed hyperedges that can be used to create the former one (see Fig.~\ref{fig:directed2undirected} for an example with $d=3$ and $m=2$).
\begin{figure}[h!]
    \centering
    \includegraphics[width=12cm ]{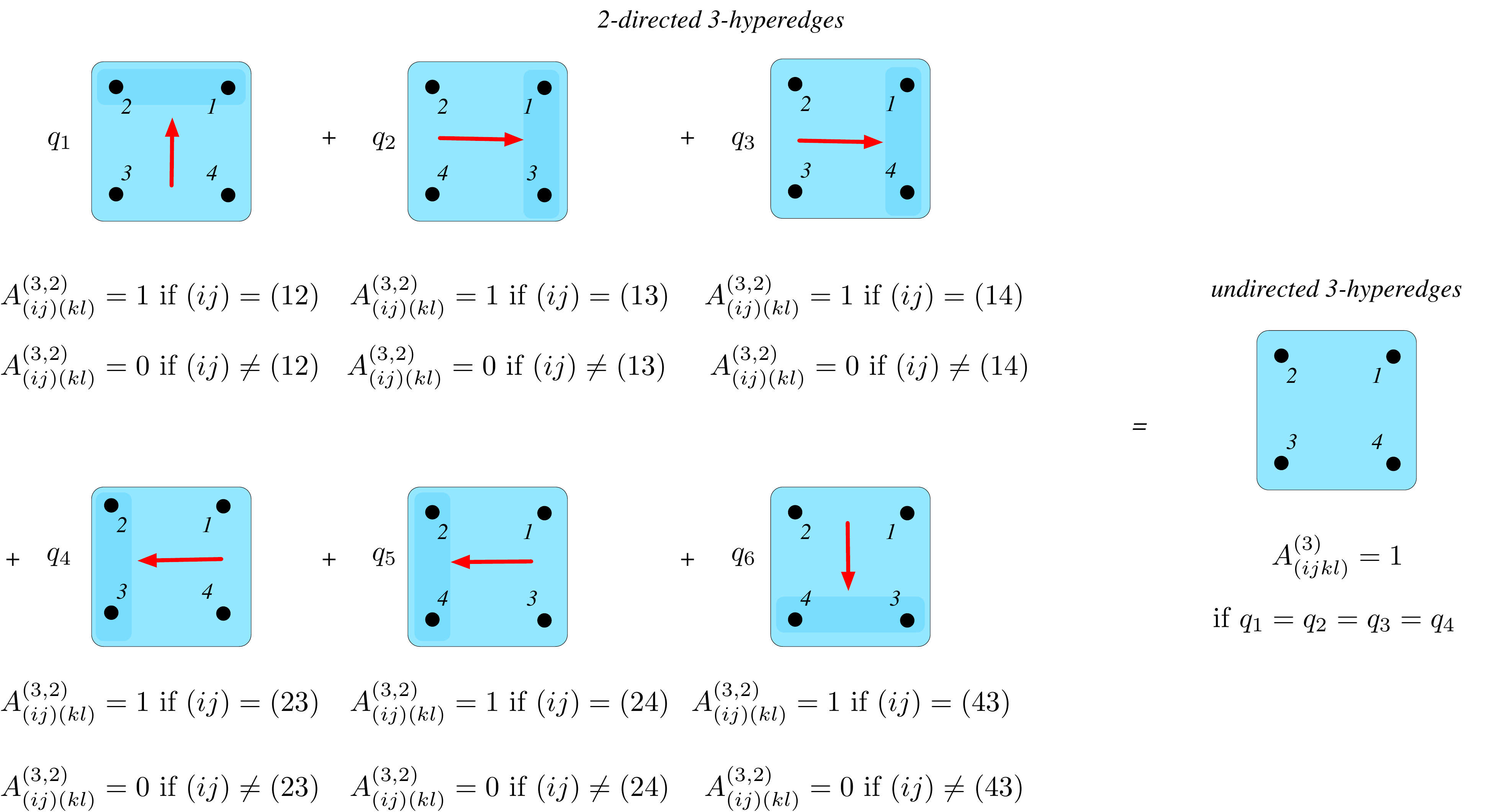}
    \caption{$m$-directed versus undirected $3$-hyperedges, the case $m=2$. The undirected $3$-hyperedge can be obtained as a ``juxtaposition'' of six $2$-directed $3$-hyperedges weighted with positive coefficients $q_1=q_2=q_3=q_4=q_5=q_6=1/6$. For other values of the coefficients $q_i$ we  obtain a family of directed weighted $3$-hyperedges. The hyperedges heads are emphasized in darker blue and the arrows help the reader to determine the directionality.}
    \label{fig:directed2undirected}
\end{figure}

By starting from the above construction for a single hyperedge, we can operate similarly to obtain a $m$-directed $d$-hypergraph as weighted sum of ``basic'' directed ones. Moreover we can act on the weights, $(q_1,q_2,q_3,q_4)$ to obtain a family of directed hypergraph, containing as particular case the undirected one (see Fig.~\ref{fig:Fig4NewRing} for the hyperring with $m=3$, $d=3$ and $Q=5$ hyperedges). The Laplace matrices $\hat{\;\mathbf{L}}^{(3,3)}$ and $\check{\;\mathbf{L}}^{(3,3)}$ can be computed by using the definitions~\eqref{eq:Lhh2} and~\eqref{eq:Lht2}, their explicit expression can be found in the Appendix~\ref{sec:appsymm1d3m}.
\begin{figure}[h!]
    \centering
    \includegraphics[width=16cm ]{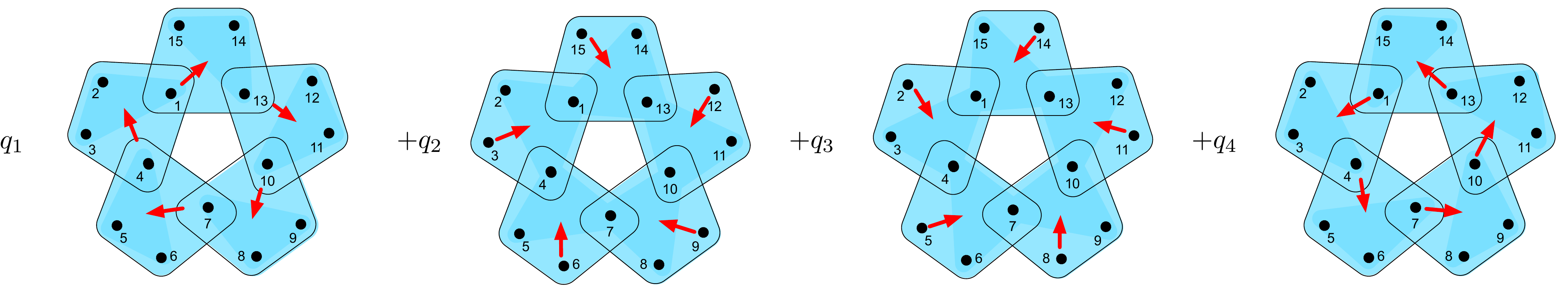}
    \caption{Weighted $3$-directed $3$-hyperring. We schematically show the family of $3$-directed $3$-hyperring built as a weighted sum of four ``basic'' $3$-directed $3$-hyperrings. The undirected $3$-hyperring can be obtained as a ``juxtaposition'' of four $3$-directed, hyperedges weighted with positive coefficients $q_1=q_2=q_3=q_4$. For the remaining values of the latter, we obtain a generic $3$-directed $3$-hyperring. The hyperedge heads are emphasized in darker blue and the arrows help the reader to determine the directionality.}
    \label{fig:Fig4NewRing}
\end{figure}

For a sake of definitiveness let us fix $q_1=1-3p$ and $q_2=q_3=q_4=p$, in this way if $p=0$ we have a $3$-directed $3$-hyperring show on the left of Fig.~\ref{fig:Fig4NewRing}. By varying $p$ we obtain a family of $3$-directed $3$-hyperring and once $p=1/4$ we get the undirected $3$-hyperring. We can thus compute for each value of $p$ the matrix $\mathbf{M}^{(3,3)}(p)$ and determine its spectrum as a function of $p$. The latter spectra are reported in the left panel of Fig.~\ref{fig:SpectrumDir2undir} for few values of $p$, as expected one can observe that for $p=1/4$ (red diamond) the spectrum is real and thus it does not have any intersection with the instability region (green region); by decreasing $p$ the spectrum becomes complex but yet it does not intersect the green region ($p=0.2$, blue square). Only smaller values of $p$ allow the spectrum to enter into the instability region ($p=0.1$ cyan triangle, and $p=0$ black circle). These findings have immediate consequences on patterns formation, indeed only $3$-directed $3$-hyperring obtained for sufficiently small values of $p$ can support Turing patterns as shown in right panel of Fig.~\ref{fig:SpectrumDir2undir}. In fact, having fixed a value of $p$ and built the $3$-directed $3$-hyperring we numerically integrate the system~\eqref{eq:manybodymDTP} and we then compute the patterns amplitude $A=\sqrt{\sum_i [(u_i(T)-u^*)^2+(v_i(T)-v^*)^2]}$ once the solutions has been stabilized, i.e., for $T$ sufficiently large. The patterns amplitude is a proxy for Turing instability, indeed if $A=0$ the solution gets back to the homogeneous state after the initial perturbation, while $A>0$ denotes the existence of a heterogeneous distribution of species.
\begin{figure}[h!]
    \centering
    \includegraphics[width=8cm ]{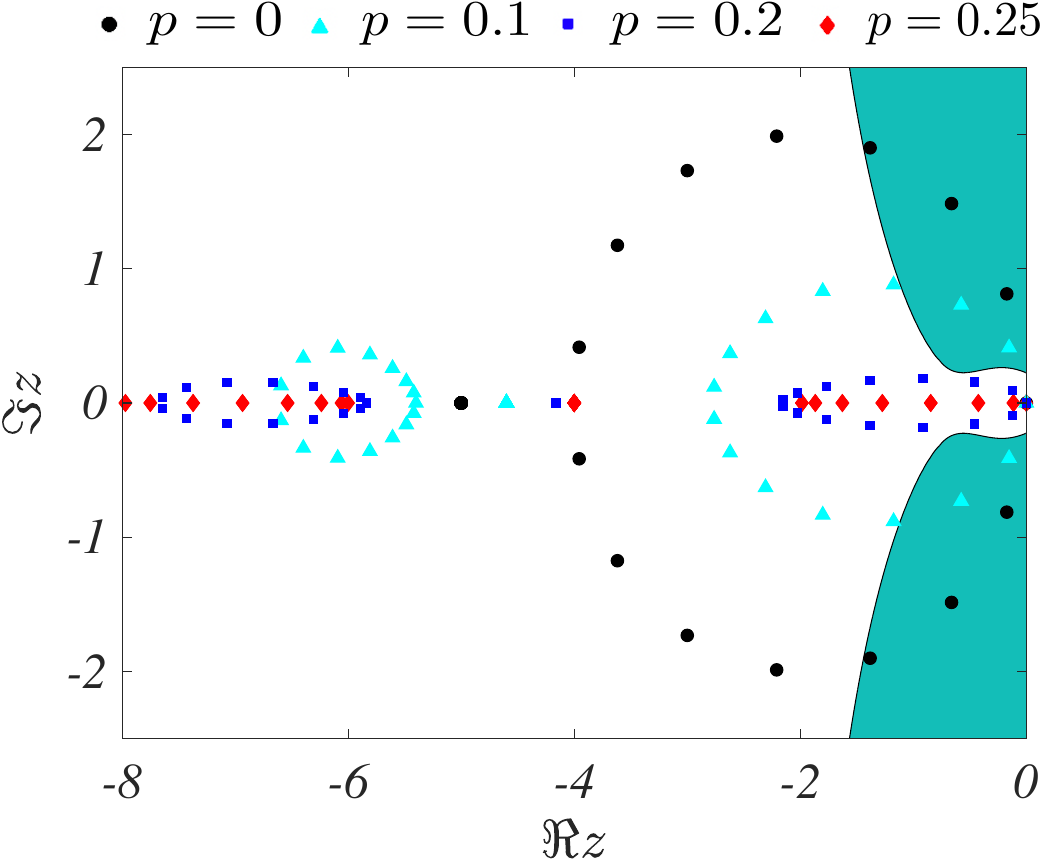}
    \includegraphics[width=8cm ]{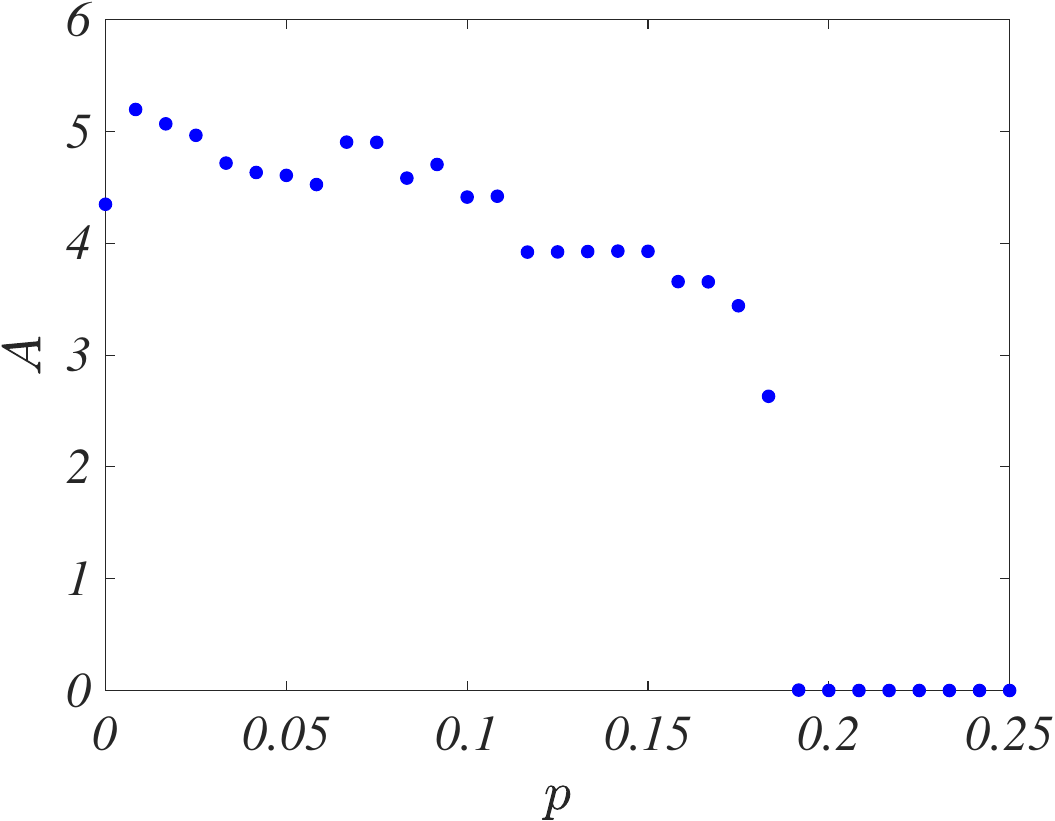}
    \caption{Spectra and pattern amplitude of the $3$-directed $3$-hyperring as a function of the parameter $p$. The left panel shows the instability region (green region) and the spectra of the matrix $\mathbf{M}^{(3,3)}(p)$ for few values of $p$ ($p=0$ black circle, $p=0.1$ cyan triangle, $p=0.2$ blue square and red diamond $p=0.25$). Right panel reports the pattern amplitude $A$ as a function of $p$, each point is the average of the patterns amplitude computed over $10$ independent simulations to reduce random effects. The remaining model parameters have been fixed to $b = 5.3$, $c = 7$. The coupling functions are the same used in Fig.~\ref{fig:d3m123Motif}, with $D_u = 1$, $D_v = 9$ and $\sigma_3=1$. }
\label{fig:SpectrumDir2undir}
\end{figure}

\section{An example of Turing patterns for $1$-directed random hypergraph}
\label{ssec:1dirrandHG}
The theory above introduced has been presented for a sake of pedagogy by adopting the framework of $m$-directed $d$-hyperring. We should however emphasize that its application domain goes far beyond this example. The aim of this section is to briefly present an application to $m$-directed random hypergraphs where directed hyperedges of different sizes are allowed for.

For a sake of definitiveness, we decide to hereby emphasize the dynamical aspect of Turing instability and thus to leave to Appendix~\ref{sec:mdirrandHG} a detailed description of the construction of the $m$-directed random hypergraph. Without loss of generality, we decide to set $m = 1$ and to consider a random hypergraph composed by $1$-directed hyperedges of sizes $3$ and $4$. The general system~\eqref{eq:manybodymDTP} rewrites thus
\begin{equation}
\label{eq:manybodymDTPrnd}
\begin{cases}
\displaystyle \frac{d{u}_i}{dt} = f_1^{(0)}\left({u}_i,{v}_i\right)&+ \sigma_2  \displaystyle \sum_{j_1,j_2}\left[h_1^{(2,1)}\left({u}_{j_1},{u}_{j_2}\right)-h_1^{(2,1)}\left({u}_{i},{u}_{i}\right)\right] A^{(2,1)}_{i(j_1j_2)}+\notag\\
&+ \sigma_3  \displaystyle \sum_{j_1,j_2,j_3}\left[h_1^{(3,1)}\left({u}_{j_1},{u}_{j_2},{u}_{j_3}\right)-h_1^{(3,1)}\left({u}_{i},{u}_{i},{u}_{i}\right)\right] A^{(3,1)}_{i(j_1\dots j_3)}\\
\displaystyle \frac{d{v}_i}{dt} = f_2^{(0)}\left({u}_i,{v}_i\right)&+ \sigma_2  \displaystyle \sum_{j_1,j_2}\left[h_2^{(2,1)}\left({v}_{j_1},{v}_{j_2}\right)-h_2^{(2,1)}\left({v}_{i},{v}_{i}\right)\right] A^{(2,1)}_{i(j_1j_2)}+\notag\\
&+ \sigma_3  \displaystyle \sum_{j_1,j_2,j_3}\left[h_2^{(3,1)}\left({v}_{j_1},{v}_{j_2},{v}_{j_3}\right)-h_2^{(3,1)}\left({v}_{i},{v}_{i},{v}_{i}\right)\right] A^{(3,1)}_{i(j_1\dots j_3)}\, .
\end{cases}
\end{equation}
Moreover we will assume the reaction terms $(f_1^{(0)},f_2^{(0)})$ to be given by the Brusselator model~\eqref{eq:brusselator}, we set $\sigma_2=\sigma_3=1$ and the coupling functions have been fixed to
\begin{equation}
    h_1^{(2,1)}=u_1^3 u_2^2\, , h_2^{(2,1)}=v_1^3 v_2^2\, , h_1^{(3,1)}=u_1^3u_2u_2\quad \text{and} \quad h_2^{(3,1)}=v_1^3v_2v_2\, ,
\end{equation}
in this way a straightforward computation starting from the definitions~\eqref{eq:hatJcheckJ} allows to provide
\begin{equation}
\label{eq:hatcheckJ}
\hat{\mathbf{J}}^{(2,1)} =\hat{\mathbf{J}}^{(3,1)} =0\quad\text{and}\quad\check{\mathbf{J}}^{(2,1)} =\check{\mathbf{J}}^{(3,1)}  = \left(
\begin{matrix}
 5(u^{*})^{4}D_u & 0\\
 0 & 5(v^{*})^{4}D_v
\end{matrix}
 \right)\, 
\end{equation}
where $(u^*,v^*)$ is the equilibrium solution of the Brusselator model. After linearization about the latter solution, we obtain
\begin{equation}
\label{eq:manybodymDlinLapCompact3_E1}
\frac{d\delta\vec{x}}{dt} = \left[\mathbb{I}_N\otimes \mathbf{J}^{(0)}+ \check{\;\mathbf{L}}^{(2,1)}\otimes \check{\mathbf{J}}^{(2,1)}+\check{\;\mathbf{L}}^{(3,1)}\otimes \check{\mathbf{J}}^{(3,1)}\right] \delta \vec{x} = \left[\mathbb{I}_N\otimes \mathbf{J}^{(0)}+ \mathbf{M}\otimes \check{\mathbf{J}}^{(2,1)}\right] \delta \vec{x} \, ,
\end{equation}
where the last equality follows by~\eqref{eq:hatcheckJ} and the definition $\mathbf{M}=\check{\;\mathbf{L}}^{(2,1)}+\check{\;\mathbf{L}}^{(3,1)}$. By computing the eigenvalues and eigenvectors of $\mathbf{M}$ one can project on the latter and by applying Eq.~\eqref{eq:instregcmplx} we can determine the instability region for the Turing instability to set. Results reported in Fig.~\ref{fig:FigRandHG} show that the $1$-directed random hypergraph supports Turing patterns, indeed there are complex eigenvalues $\Lambda^{(s)}$ laying inside the instability region, determining thus a positive dispersion relation and the emergence of patterns. Let us observe that the instability region does not intersect the real axis and thus a symmetrical version of the hypergraph would not exhibit patterns, being its spectrum real.
\begin{figure}[h!]
    \centering
    \includegraphics[width=18cm ]{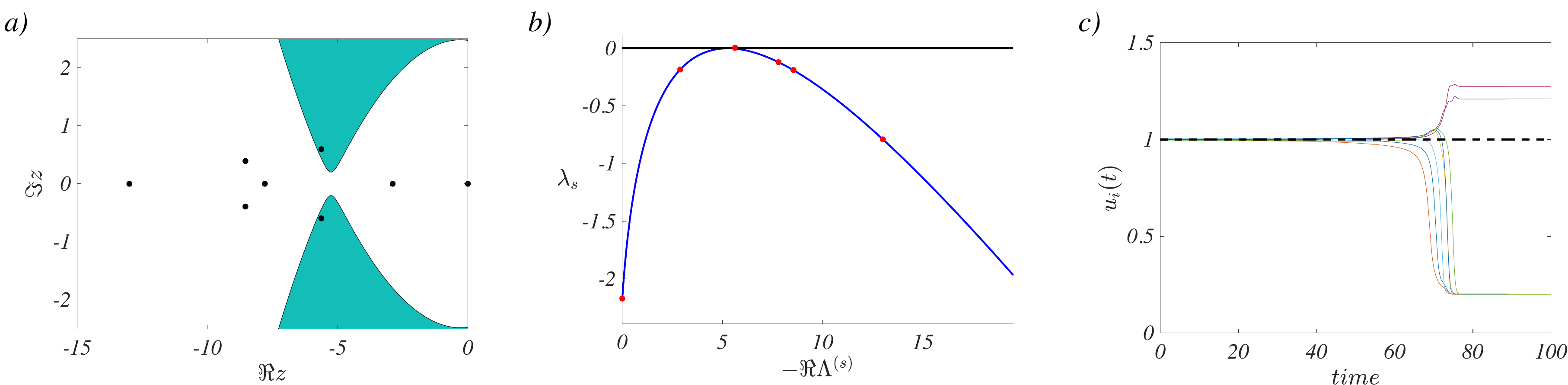}
    \caption{$1$-directed random hypergraph. Panel a): the instability region (green region) and the complex spectrum $\Lambda^{(s)}$ (black dots) of the matrix $\mathbf{M}$ (see text); we can appreciate the presence of two complex conjugated eigenvalues inside the instability region determining thus the onset of Turing instability. Panel b): the dispersion relation, we can observe the presence of an eigenvalue for which the dispersion relation is positive (red dots), signalizing again the emergence of Turing patterns as one can see in panel c) where we report $u_i(t)$ versus time. The underlying $1$-directed random hypergraph has $8$ nodes shared among $20$ $1$-directed hyperedges, $17$ of which have size $3$ and the remaining three have size $4$. The construction of the random hypergraph can be found in Appendix~\ref{sec:mdirrandHG}. The local dynamics is given by the Brusselator with parameters $b=5.3$, $c=12$, the coupling is realized with $\sigma_d=1$ for $d=2,3$ and functions $h_1^{(2,1)}=D_uu_1^3 u_2^2$, $h_2^{(2,1)}=D_vv_1^3 v_2^2$, $h_1^{(3,1)}=D_uu_1^3u_2u_2$ and $h_2^{(3,1)}=D_vv_1^3v_2v_2$, where $D_u=0.0495$ and $D_v=9.2$.}
    \label{fig:FigRandHG}
\end{figure}

\begin{remark}
Before to conclude, let us emphasize a peculiar result holding true for $m$-directed $d$-hyperring. As shown in Fig.~\ref{fig:d2m12Motif} and Fig.~\ref{fig:d3m123Motif}, the spectrum ``moves'' toward the imaginary axis as $m$ increases. This implies that if we can find model parameters, $b,c$, diffusive constants $D_u,D_v$ and coupling functions $h_i^{(d,m)}$, $i=1,2$ for which the instability region is ``far enough'' from the imaginary axis, then Turing instability could emerge more easily for small $m$ than for larger ones. A numerical confirmation of this claim is reported in the left panel of Fig.~\ref{fig:TPforsmallm} where we show the instability region (green region) together with the spectrum of $1$-directed $3$-hyperring (black dots) together with the one for the $3$-directed $3$-hyperring (blue squares), both hyperring be formed by $Q=15$ hyperedges. One can observe the presence of black dots inside the instability region, while no blue square enters into it, hence Turing patterns cannot develop on the latter while they are present in the former. The dynamical system used to present this result is the Brusselator with parameters, $b=5.4$, $c=7$; the coupling functions are $h^{(3,1)}_1=D_uu_1^2u_2u_3/2$, $h^{(3,1)}_2=D_vv_1^2v_2v_3/2$, $h^{(3,3)}_1=D_uu_1u_2u_3^2$ and $h^{(3,3)}_2=D_vv_1v_2v_3^2$, where $D_u=0.015$ and $D_v=0.31$. The right panel of Fig.~\ref{fig:TPforsmallm} shows the combinations of $D_u$ and $D_v$, having fixed all the above parameters and functions but  the two diffusion coefficients, that produce this phenomenon. We can appreciate (see the inbox) that only $D_u$ and $D_v$ belonging to a very tiny region of the plane colored in dark blue (patterns for $m=1$ and $m=2$) and cyan (patterns for $m=1$), return this result. We can hence conclude that this phenomenom is very rare.
\begin{figure}[h!]
    \centering
    \includegraphics[width=9cm ]{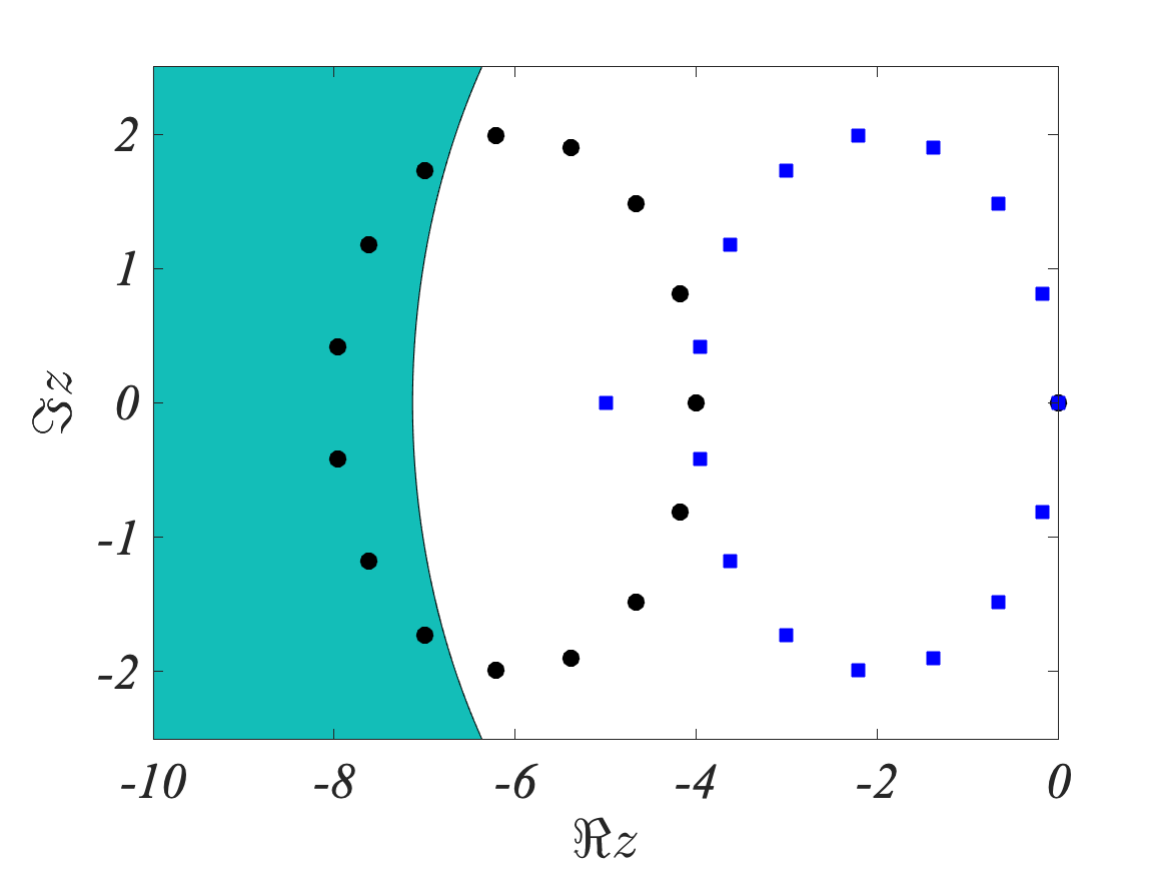}\quad     \includegraphics[width=8cm ]{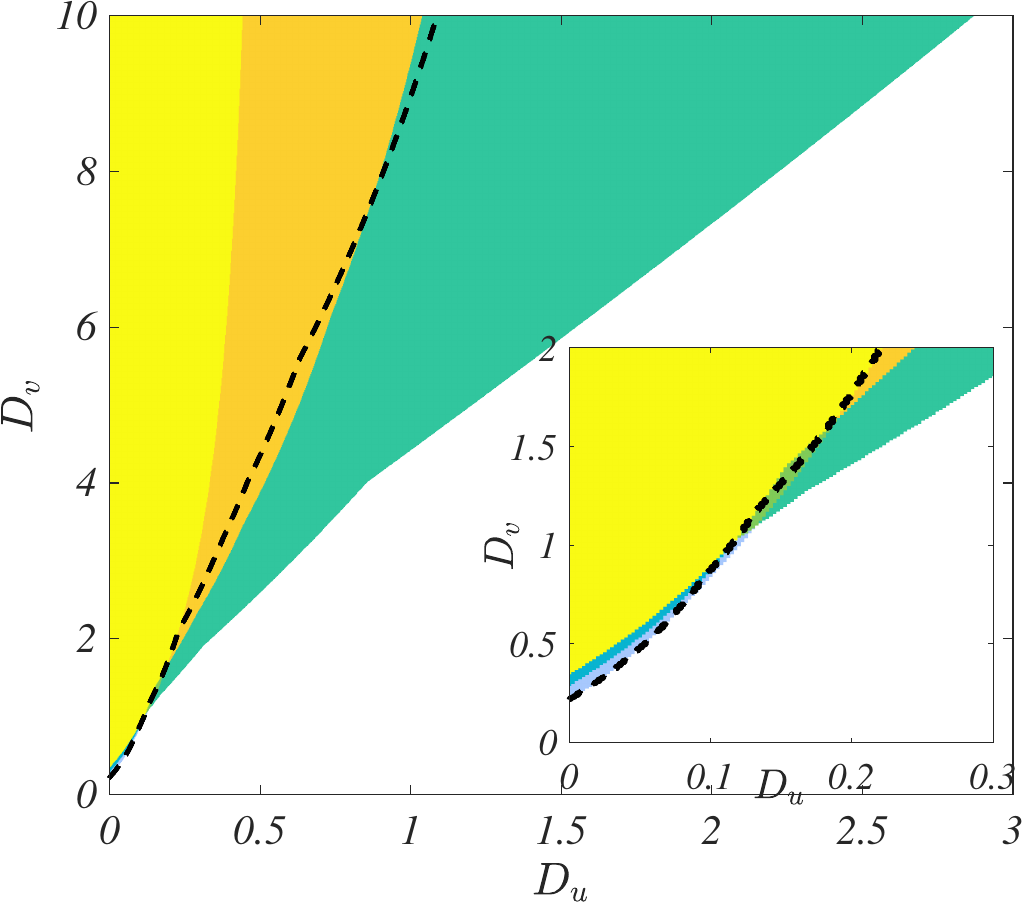}
    \caption{Left panel: instability region (green area) and spectra of $m$-directed $3$-hyperring, $m=1$ (black dots) and $m=3$ (blue squares). Right panel: emergence of patterns for low directionality in the plane $D_u$, $D_v$. Parameters in the yellow region return patterns for $1$, $2$ and $3$-directed hyperring, the orange region denotes presence of patterns for $2$ and $3$-directed hyperring, while the green one only for $3$-directed hyperring. The two small blue regions are associated to the presence of patterns for low directionality, more precisely in the dark blue region patterns emerge for $m=1$ and $m=2$, while cyan region is associated to $m=1$. The region bounded by the black curve is associated to patterns for the undirected hyperring.}
    \label{fig:TPforsmallm}
\end{figure}
Let us observe the peculiar shape of the instability region that has a non-empty intersection with the negative real axis, implying thus that also a symmetrical $3$-hyperring will support Turing patterns. Eventually, we would like to stress that an analogous phenomenon is also possible for directed network.
\end{remark}

\section{Conclusions}
\label{sec:conc}
In this paper, we provided a contribute to the literature on Turing patterns by investigating their emergence in directed higher-order topologies. We present a novel model for dynamical processes on directed hypergraphs, where each hyperedge has a tail and a head, with tail nodes pointing towards head nodes, the state of the latter being thus influenced by the former. By utilizing this framework, we extend and generalize Turing's theory to directed hypergraphs. For simplicity and illustrative purposes, we use a simplified version of the proposed theoretical framework, specifically by focusing on $m$-directed $d$-hypergraphs and the Brusselator, a paradigmatic model in the framework of Turing patterns.
Our findings indicate that increased directionality (higher $m$, implying more nodes in the head of the hypergraph) facilitates the emergence of patterns. This observation extends existing knowledge on graphs, demonstrating that directionality enhances patterns formation. Through parameter analysis, we identify specific conditions under which directionality may inhibit patterns emergence.
To compare patterns between directed and undirected hypergraphs, we detail the methodology used to transform directed hypergraphs into symmetric ones while preserving hyperdegree distribution. It is important to note that the simplifications made in this paper serve illustrative purposes, and our results are applicable to more general directed higher-order structures. We believe that our results provide thus a strong contribution to the active framework of higher-order systems.

\noindent
\\
{\bf Acknowledgments}\\
R.M. acknowledges JSPS, Japan KAKENHI JP22K11919, JP22H00516, and JST, Japan CREST JP-MJCR1913
for financial support. T.C. and R.M. would like to thank Luca Gallo for interesting discussions had in a preliminary phase of this project about $m$-directed hypergraph.

\bibliographystyle{abbrv}
\bibliography{sample}

\appendix

\section{Generalized natural coupling: proof in the general setting}
\label{sec:gencaseHcond}

In the main text we have introduced a working assumption allowing us to factorize the terms involving the derivatives of the coupling functions and thus to pass from Eq.~\eqref{eq:manybodymDlinLapCompact} to Eq.~\eqref{eq:manybodymDlinLapCompact} that is the starting point of the construction of the dispersion relation and the inequality {defining} the instability region.

The goal of this section is to consider the general hypothesis for which Eq.~\eqref{eq:condJJ} holds true and thus the same conclusions drawn in the main text can be applied here. Assume thus each component of the coupling function $\vec{h}^{(d,m)}=(h_1^{(d,m)},\dots,h_n^{(d,m)})^\top$ to be a monomial, namely
\begin{equation}
\label{eq:gnch}
h^{(d,m)}_i(\vec{x}_1,\dots,\vec{x}_d)=C_{d}^{(i)}\vec{x}_1^{\vec{a}^{(i)}_1}\dots \vec{x}_d^{\vec{a}^{(i)}_d}\, ,
\end{equation}
for some real multi-index ${\vec{a}^{(i)}_1},\dots ,{\vec{a}^{(i)}_d}$, i.e., $\vec{a}^{(i)}_\ell=({a}^{(i)}_{\ell,1},\dots,{a}^{(i)}_{\ell,n})^\top$, a real coefficient $C_{d}^{(i)}$ and we used the notation $\vec{x}_\ell^{\vec{a}^{(i)}_\ell}:={x}_{\ell,1}^{{a}^{(i)}_{\ell,1}}\dots {x}_{\ell,n}^{{a}^{(i)}_{\ell,n}}$. Assume moreover a condition similar to~\eqref{eq:conditionsimple} to be satisfied, namely for all $j=1,\dots,n$
\begin{equation}
\label{eq:conditionsimplej}
 \alpha^{(d,m)} = \frac{a_{1,j}^{(i)}+\dots +a^{(i)}_{m-1,j}}{a_{m,j}^{(i)}+\dots + a^{(i)}_{d,j}}\, .
\end{equation}
Then we can conclude that 
\begin{equation}
\label{eq:condJhatJcheck}
\hat{\mathbf{J}}^{(d,m)}=\alpha^{(d,m)}\check{\mathbf{J}}^{(d,m)}\, .
\end{equation}

By using the definition of the latter matrices and the functional form for $h_i^{(d,m)}$, we get for all $i,j=1,\dots,n$
\begin{align*}
\hat{J}_{ij}^{(d,m)}=\frac{\partial h_i^{(d,m)}}{\partial x_{1,j}}+\dots+ \frac{\partial h_i^{(d,m)}}{\partial x_{m-1,j}}&=C_{d}^{(i)}(x_j^*)^{a^{(i)}_{1,j}+\dots a^{(i)}_{d,j}-1}(a^{(i)}_{1,j}+\dots +a^{(i)}_{m-1,j})\displaystyle\Pi_{\substack{k=1,\dots,n \\k\neq j}} (x_k^*)^{a^{(i)}_{1,k}+\dots +a^{(i)}_{d,k}}\text{ and } \\
\check{J}_{ij}^{(d,m)}=\frac{\partial h_i^{(d,m)}}{\partial x_{m,j}}+\dots+ \frac{\partial h_i^{(d,m)}}{\partial x_{d,j}}&=C_{d}^{(i)}(x_j^*)^{a^{(i)}_{1,j}+\dots +a^{(i)}_{d,j}-1}(a^{(i)}_{m,j}+\dots a^{(i)}_{d,j})\displaystyle\Pi_{\substack{k=1,\dots,n \\k\neq j}} (x_k^*)^{a^{(i)}_{1,k}+\dots +a^{(i)}_{d,k}}\, ,
\end{align*}
by invoking~\eqref{eq:conditionsimplej} we can rewrite the previous equation as
\begin{eqnarray*}
\hat{J}_{ij}^{(d,m)}=\frac{\partial h_i^{(d,m)}}{\partial x_{1,j}}+\dots+ \frac{\partial h_i^{(d,m)}}{\partial x_{m-1,j}}=\alpha^{(d,m)} \left(\frac{\partial h_i^{(d,m)}}{\partial x_{m,j}}+\dots+ \frac{\partial h_i^{(d,m)}}{\partial x_{d,j}}\right)=\alpha^{(d,m)}\check{J}_{ij}^{(d,m)}\, .
\end{eqnarray*}

\section{Details for the computation of the instability region in the complex plane}
\label{sec:reldispcmplx}

In this section, we detail the computation needed to determine the instability region~\eqref{eq:instregcmplx} for which the system exhibits Turing patterns. One can come back to Eq.~\eqref{eq:RelDisp} and notice that because we work with an asymmetric Laplace matrix, its eigenvalues can be complex. We can then rewrite $\Lambda^{(s)}=\Re(\Lambda^{(s)})+ \iota\Im(\Lambda^{(s)})$, where $\iota=\sqrt{-1}$ and define
\begin{eqnarray*}
 \Re[\mathrm{tr}(\textbf{J}_s)]&=& \mathrm{tr}(\mathbf{J}_0)+\sigma_d\mathrm{tr}(\check{\mathbf{J}}^{(d,m)})\Re\Lambda^{(s)}\\
 \Im[\mathrm{tr}(\textbf{J}_s)]&=& \sigma_d\mathrm{tr}(\check{\mathbf{J}}^{(d,m)})\Im\Lambda^{(s)}\\ 
\Re[\det(\textbf{J}_s)]&=&\det(\mathbf{J}_0)+\sigma_d\left((\mathbf{J}_0)_{11}(\check{\mathbf{J}}^{(d,m)})_{22}+(\mathbf{J}_0)_{22} (\check{\mathbf{J}}^{(d,m)})_{11}\right)\Re\Lambda^{(s)}+\sigma_d^2 (\check{\mathbf{J}}^{(d,m)})_{11}(\check{\mathbf{J}}^{(d,m)})_{22}\left[(\Re\Lambda^{(s)})^2-(\Im\Lambda^{(s)})^2\right]\\
 \Im[\det(\textbf{J}_s)]&=& \sigma_d\left((\mathbf{J}_0)_{11}(\check{\mathbf{J}}^{(d,m)})_{22}+(\mathbf{J}_0)_{22} (\check{\mathbf{J}}^{(d,m)})_{11}\right)\Im\Lambda^{(s)}+2\sigma_d^2 (\check{\mathbf{J}}^{(d,m)})_{11}(\check{\mathbf{J}}^{(d,m)})_{22}\Re\Lambda^{(s)}\Im\Lambda^{(s)}\, ,
\end{eqnarray*}
and thus rewrite the real part of the dispersion relation as 
\begin{equation*}
    \Re(\lambda_s)=\frac{1}{2}\left[ \Re\mathrm{tr}(\textbf{J}_s) + \gamma \right]\, ,
\end{equation*}
where $\displaystyle\gamma=\sqrt{\frac{A+\sqrt{A^2+B^2}}{2}}$ and $A=\left(\Re[\mathrm{tr}(\textbf{J}_s)]\right)^2-\left(\Im[\mathrm{tr}(\textbf{J}_s)]\right)^2-4\Re[\det(\textbf{J}_s)]$ and $ B=2 \Re[\mathrm{tr}(\textbf{J}_s)] \Im[\mathrm{tr}(\textbf{J}_s)]-4\Im[\det(\textbf{J}_s)]$. The condition to obtain Turing instability is to have some $s$ such that $\Re(\lambda_s)>0$, i.e., $\Re[\mathrm{tr}(\textbf{J}_s)]>-\gamma$, which can be rewritten as
\begin{equation*}
    S_2\left( \Re\Lambda^{(s)}\right)\left(\Im\Lambda^{(s)}\right)^2< -S_1\left(\Re\Lambda^{(s)}\right)\, ,
\end{equation*}
where $S_1$ and $S_2$ are polynomials explicitly given by
\begin{eqnarray*}
    S_1\left( x\right) &=&C_{14}x^4+C_{13}x^3+C_{12}x^2+C_{11}x+C_{10}\\
    S_2\left( x\right) &=&C_{22}x^2+C_{21}x+C_{20}\, ,
\end{eqnarray*}
and where the coefficients $C_{ij}$ are 
\begin{eqnarray*}
C_{14} &=&\sigma_d^4 (\check{\mathbf{J}}^{(d,m)})_{11} (\check{\mathbf{J}}^{(d,m)})_{22}\left[(\check{\mathbf{J}}^{(d,m)})_{11}+ (\check{\mathbf{J}}^{(d,m)})_{22}\right]^2\\
C_{13} &=&\sigma_d^3\left[(\check{\mathbf{J}}^{(d,m)})_{11}+ (\check{\mathbf{J}}^{(d,m)})_{22}\right]^2\left[ (\mathbf{J}_0)_{11} (\check{\mathbf{J}}^{(d,m)})_{22}+(\mathbf{J}_0)_{22} (\check{\mathbf{J}}^{(d,m)})_{11}\right]+\\
 &+&2\sigma_d^3\mathrm{tr}(\mathbf{J}_0) (\check{\mathbf{J}}^{(d,m)})_{11} (\check{\mathbf{J}}^{(d,m)})_{22}\left[ (\check{\mathbf{J}}^{(d,m)})_{11}+ (\check{\mathbf{J}}^{(d,m)})_{22}\right]\\
C_{12} &=&\sigma_d^2\det(\mathbf{J}_0)( (\check{\mathbf{J}}^{(d,m)})_{11}+ (\check{\mathbf{J}}^{(d,m)})_{22})^2+\sigma_d^2\left(\mathrm{tr}(\mathbf{J}_0)\right)^2 (\check{\mathbf{J}}^{(d,m)})_{11} (\check{\mathbf{J}}^{(d,m)})_{22}+\\
&+&2\sigma_d^2\mathrm{tr}(\mathbf{J}_0)( (\check{\mathbf{J}}^{(d,m)})_{11}+ (\check{\mathbf{J}}^{(d,m)})_{22})\left[ (\mathbf{J}_0)_{11} (\check{\mathbf{J}}^{(d,m)})_{22}+(\mathbf{J}_0)_{22} (\check{\mathbf{J}}^{(d,m)})_{11}\right]\\
C_{11} &=&2\sigma_d^2\mathrm{tr}(\mathbf{J}_0)( (\check{\mathbf{J}}^{(d,m)})_{11}+ (\check{\mathbf{J}}^{(d,m)})_{22})\det(\mathbf{J}_0)+\left(\mathrm{tr}(\mathbf{J}_0)\right)^2\sigma_d^2\left[ (\mathbf{J}_0)_{11} (\check{\mathbf{J}}^{(d,m)})_{22}+(\mathbf{J}_0)_{22} (\check{\mathbf{J}}^{(d,m)})_{11}\right]\\
C_{10}&=&\det(\mathbf{J}_0)\left(\mathrm{tr}(\mathbf{J}_0)\right)^2\\
C_{22} &=&\sigma_d^4 (\check{\mathbf{J}}^{(d,m)})_{11} (\check{\mathbf{J}}^{(d,m)})_{22}\left[(\check{\mathbf{J}}^{(d,m)})_{11}- (\check{\mathbf{J}}^{(d,m)})_{22}\right]^2\\
C_{21} &=&\sigma_d^3\left[ (\mathbf{J}_0)_{11} (\check{\mathbf{J}}^{(d,m)})_{22}+(\mathbf{J}_0)_{22}(\check{\mathbf{J}}^{(d,m)})_{11}\right]\left[ (\check{\mathbf{J}}^{(d,m)})_{11}- (\check{\mathbf{J}}^{(d,m)})_{22}\right]^2\\
C_{20}&=&\sigma_d^2(\mathbf{J}_0)_{11}(\mathbf{J}_0)_{22}\left[ (\check{\mathbf{J}}^{(d,m)})_{11}-(\check{\mathbf{J}}^{(d,m)})_{22}\right]^2\, .
\end{eqnarray*}

\section{Another example of Turing patterns on a $m$-directed $3$-hyperring}
\label{sec:anotherexTP}
This section is an extension of Section~\ref{sec:TP}, whose aim is to show another example of patterns emergence in $m$-directed $d$-hyperring. As in Section~\ref{sec:TP}, we consider a $m$-directed $3$-hyperring of $Q=15$ hyperedges, having thus $N=45$ and the Brusselator model as dynamical system, with respect to Section~\ref{sec:TP} the parameters model have now been set to $b=5.5$ and $c=7$ in such a way the instability region in the complex plane forms a sort of vertical band that include the imaginary axis minus a small neighbor of the origin. This implies that also symmetrical $d$-hyperrings could support Turing patterns. The coupling function used are also the same of the previous Section, namely
\begin{equation*}
 \vec{h}^{(3,1)}=\frac{1}{2}\left(D_u u_{\ell_1}^2u_{\ell_2}u_{\ell_3},D_v v_{\ell_1}^2v_{\ell_2}v_{\ell_3}\right)^\top\, , \vec{h}^{(3,2)}=\left(D_u u_{\ell_1}^2u_{\ell_2}u_{\ell_3},D_v v_{\ell_1}^2v_{\ell_2}v_{\ell_3}\right)^\top\text{ and } \vec{h}^{(3,3)}=\left(D_u u_{\ell_1}u_{\ell_2}u^2_{\ell_3},D_v v_{\ell_1}v_{\ell_2}v^2_{\ell_3}\right)^\top\, ,
\end{equation*}
in such a way to also obtain
\begin{equation*}
\hat{\mathbf{J}}^{(3,1)} = 0\,, \check{\mathbf{J}}^{(3,1)} = \check{\mathbf{J}}^{(3,2)}=\check{\mathbf{J}}^{(3,3)}=\hat{\mathbf{J}}^{(3,2)}=\hat{\mathbf{J}}^{(3,3)}=\displaystyle
\left(\begin{matrix}
 2D_u(u^*)^2 &  0\\
  0 &   2D_v(v^*)^2
\end{matrix}\right)\, .
\end{equation*}

First of all, let us observe that Turing patterns emerge in the symmetrical $3$-hyperring (see top row panels in Fig.~\ref{fig:d3m123Motifbis}), indeed the spectrum (black dots) enters into the instability region (panel $a_1)$). Then in the case $m=1$ (second row panels from the top) one can observe the finite size effect; indeed the dispersion relation is positive (blue line in panel $b_2)$, however, because of the discrete nature of the Laplace spectrum, it happens that no eigenvalues lie in this region and thus patterns cannot develop (see panel $c_2)$. In the cases $m=2$ (third row panels from the top) and $m=3$ (last row panels from the bottom) Turing patterns emerge because the spectrum falls into the instability region.
\begin{figure}[h!]
    \centering
    \includegraphics[width=18cm ]{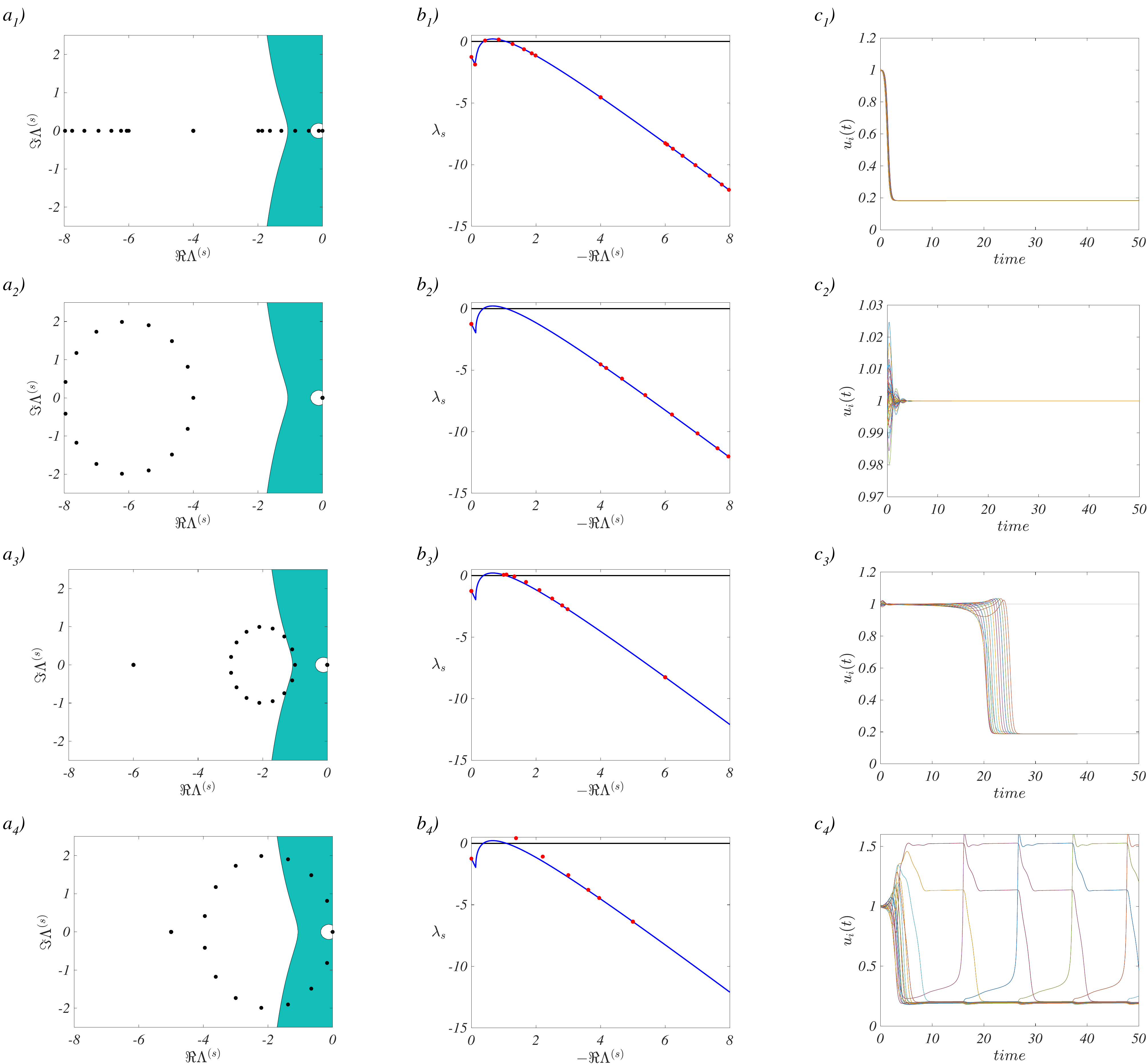}
    \caption{Undirected (top row panels) and $m$-directed $3$-hyperring, $m=1$ (second row panels from the top), $m=2$ (third row panels from the top) and $m=3$ (bottom row panels). Panels $a_1)$, $a_2)$, $a_3)$ and $a_4)$ present the instability region in the complex plane (green area) and the complex spectrum of the effective Laplace matrix $\mathbf{M}^{(d,m)}$ (black dots). Panels $b_1)$, $b_2)$, $b_3)$ and $b_4)$ report the dispersion relation $\lambda_s$ as a function of $-\Re\Lambda^{(s)}$. The time evolution of $u_i(t)$ is presented in panels $c1)$, $c_2)$, $c_3)$ and $c_4)$. The parameter of the Brusselator model are given by $b=5.5$, $c=7$, $D_u=1$, $D_v=9$ and the coupling functions are given by $h_1^{(3,1)}(u_1,u_2,u_3)=\frac{D_u}{2}u_1^2u_2u_3$,  $h_2^{(3,1)}(v_1,v_2,v_3)=\frac{D_v}{2}v_1^2v_2v_3$, $h_1^{(3,2)}(u_1,u_2,u_3)=D_u u_1^2u_2u_3$, $h_2^{(3,2)}(v_1,v_2,v_3)=D_v v_1^2v_2v_3$,  $h_1^{(3,3)}(u_1,u_2,u_3)=D_u u_1u_2u_3^2$ and $h_2^{(3,3)}(v_1,v_2,v_3)=D_v v_1v_2v_3^2$.}
    \label{fig:d3m123Motifbis}
\end{figure}

\section{More details about the construction of the family of weighted directed hyperring and the associated Laplace matrices.}
\label{sec:appsymm1d3m}

The aim of this section is to present the details for the construction of the matrices $\hat{\;\mathbf{L}}^{(3,3)}$ and $\check{\;\mathbf{L}}^{(3,3)}$ for the weighted $3$-directed $3$-hyperring. Let us refer to the hyperring on the left side of Fig.~\ref{fig:Fig4NewRing} and in particular to its hyperedge $[1,2,3,4]$ where $[1,2,3]$ are in the head and $4$ in the tail; the analysis for the other ones will be similar. By using the definitions~\eqref{eq:Lhh2} and~\eqref{eq:Lht2} (or equivalently~\eqref{eq:Lhh} and~\eqref{eq:Lht}) we straightforwardly get its contribution to $\hat{\;\mathbf{L}}^{(3,3)}$ to be given by
\begin{equation*}
q_1\left(
\begin{matrix}
 -2 & 1 & 1 & 0\\
 1 & -2 & 1 & 0\\
  1 & 1 & -2 & 0\\
   0 & 0 & 0 & 0
\end{matrix}\right)\, ,
\end{equation*}
while for $\check{\;\mathbf{L}}^{(3,3)}$
\begin{equation*}
q_1\left(
\begin{matrix}
 -2 & 0 & 0 & 2\\
 0 & -2 & 0 & 2\\
  0 & 0 & -2 & 0\\
   0 & 0 & 0 & -2
\end{matrix}\right)\, ,
\end{equation*}
let us observe that node $4$ is in the head of the hyperedge $[4,5,6,7]$, receiving $2$ contribution from node $7$ that explain the $-2$ element in the position $(4,4)$ of the latter matrix.

Let us now consider the second from the left hyperring of Fig.~\ref{fig:Fig4NewRing} and in particular to its hyperedge $[1,2,3,4]$ where $[1,2,4]$ are in the head and $3$ in the tail. Reasoning as before we get its contribution to $\hat{\;\mathbf{L}}^{(3,3)}$ to be
\begin{equation*}
q_2\left(
\begin{matrix}
 -4 & 1 & 0 & 1\\
 1 & -2 & 1 & 0\\
   0 & 0 & 0 & 0\\
  1 & 1 & 0 & -4
\end{matrix}\right)\, ,
\end{equation*}
let us observe that node $1$ is also part of the hyperedge $[1,13,14,15]$ from which it receives two contribution, equalizing thus the $-4$ in position $(1,1)$. Similarly for node $4$ belonging to $[4,5,6,7]$.
For $\check{\;\mathbf{L}}^{(3,3)}$ we get
\begin{equation*}
q_2\left(
\begin{matrix}
 -4 & 0 & 2 & 0\\
 0 & -2 & 2 & 0\\
  0 & 0 & 0 & 0\\
   0 & 0 & 2 & -4
\end{matrix}\right)\, .
\end{equation*}

By considering the remaining hyperedges and using similar ideas one can obtain the following expressions for the Laplace matrices
\begin{equation*}
{\tiny\hat{\;\mathbf{L}}^{(3,3)}=\left(\begin{array}{ccccccccccccccc} -2\ell_{11} & q_{2}+q_{1} & q_{3}+q_{1} & q_{3}+q_{2} & 0 & 0 & 0 & 0 & 0 & 0 & 0 & 0 & q_{3}+q_{2} & q_{4}+q_{2} & q_{4}+q_{3}\\ q_{2}+q_{1} & -2\ell_{22} & q_{4}+q_{1} & q_{4}+q_{2} & 0 & 0 & 0 & 0 & 0 & 0 & 0 & 0 & 0 & 0 & 0\\ q_{3}+q_{1} & q_{4}+q_{1} & -2\ell_{33} & q_{4}+q_{3} & 0 & 0 & 0 & 0 & 0 & 0 & 0 & 0 & 0 & 0 & 0\\ q_{3}+q_{2} & q_{4}+q_{2} & q_{4}+q_{3} & -2\ell_{11} & q_{2}+q_{1} & q_{3}+q_{1} & q_{3}+q_{2} & 0 & 0 & 0 & 0 & 0 & 0 & 0 & 0\\ 0 & 0 & 0 & q_{2}+q_{1} & -2\ell_{22} & q_{4}+q_{1} & q_{4}+q_{2} & 0 & 0 & 0 & 0 & 0 & 0 & 0 & 0\\ 0 & 0 & 0 & q_{3}+q_{1} & q_{4}+q_{1} & -2\ell_{33} & q_{4}+q_{3} & 0 & 0 & 0 & 0 & 0 & 0 & 0 & 0\\ 0 & 0 & 0 & q_{3}+q_{2} & q_{4}+q_{2} & q_{4}+q_{3} & -2\ell_{11} & q_{2}+q_{1} & q_{3}+q_{1} & q_{3}+q_{2} & 0 & 0 & 0 & 0 & 0\\ 0 & 0 & 0 & 0 & 0 & 0 & q_{2}+q_{1} & -2\ell_{22}& q_{4}+q_{1} & q_{4}+q_{2} & 0 & 0 & 0 & 0 & 0\\ 0 & 0 & 0 & 0 & 0 & 0 & q_{3}+q_{1} & q_{4}+q_{1} & -2\ell_{33} & q_{4}+q_{3} & 0 & 0 & 0 & 0 & 0\\ 0 & 0 & 0 & 0 & 0 & 0 & q_{3}+q_{2} & q_{4}+q_{2} & q_{4}+q_{3} & -2\ell_{11}& q_{2}+q_{1} & q_{3}+q_{1} & q_{3}+q_{2} & 0 & 0\\ 0 & 0 & 0 & 0 & 0 & 0 & 0 & 0 & 0 & q_{2}+q_{1} & -2\ell_{22} & q_{4}+q_{1} & q_{4}+q_{2} & 0 & 0\\ 0 & 0 & 0 & 0 & 0 & 0 & 0 & 0 & 0 & q_{3}+q_{1} & q_{4}+q_{1} & -2\ell_{33} & q_{4}+q_{3} & 0 & 0\\ q_{3}+q_{2} & 0 & 0 & 0 & 0 & 0 & 0 & 0 & 0 & q_{3}+q_{2} & q_{4}+q_{2} & q_{4}+q_{3} & -2\ell_{11} & q_{2}+q_{1} & q_{3}+q_{1}\\ q_{4}+q_{2} & 0 & 0 & 0 & 0 & 0 & 0 & 0 & 0 & 0 & 0 & 0 & q_{2}+q_{1} & -2\ell_{22}& q_{4}+q_{1}\\ q_{4}+q_{3} & 0 & 0 & 0 & 0 & 0 & 0 & 0 & 0 & 0 & 0 & 0 & q_{3}+q_{1} & q_{4}+q_{1} & -2\ell_{33}\end{array}\right)}\, ,
\end{equation*}
and
\begin{equation*}
 {\tiny  \check{\;\mathbf{L}}^{(3,3)}=\left(\begin{array}{ccccccccccccccc} -2\ell_{11} & 2\,q_{3} & 2\,q_{2} & 2\,q_{1} & 0 & 0 & 0 & 0 & 0 & 0 & 0 & 0 & 2\,q_{4} & 2\,q_{3} & 2\,q_{2}\\ 2\,q_{4} & -2\ell_{22} & 2\,q_{2} & 2\,q_{1} & 0 & 0 & 0 & 0 & 0 & 0 & 0 & 0 & 0 & 0 & 0\\ 2\,q_{4} & 2\,q_{3} & -2\ell_{33} & 2\,q_{1} & 0 & 0 & 0 & 0 & 0 & 0 & 0 & 0 & 0 & 0 & 0\\ 2\,q_{4} & 2\,q_{3} & 2\,q_{2} & -2\ell_{11} & 2\,q_{3} & 2\,q_{2} & 2\,q_{1} & 0 & 0 & 0 & 0 & 0 & 0 & 0 & 0\\ 0 & 0 & 0 & 2\,q_{4} & -2\ell_{22}& 2\,q_{2} & 2\,q_{1} & 0 & 0 & 0 & 0 & 0 & 0 & 0 & 0\\ 0 & 0 & 0 & 2\,q_{4} & 2\,q_{3} & -2\ell_{33}& 2\,q_{1} & 0 & 0 & 0 & 0 & 0 & 0 & 0 & 0\\ 0 & 0 & 0 & 2\,q_{4} & 2\,q_{3} & 2\,q_{2} & -2\ell_{11}& 2\,q_{3} & 2\,q_{2} & 2\,q_{1} & 0 & 0 & 0 & 0 & 0\\ 0 & 0 & 0 & 0 & 0 & 0 & 2\,q_{4} & -2\ell_{22} & 2\,q_{2} & 2\,q_{1} & 0 & 0 & 0 & 0 & 0\\ 0 & 0 & 0 & 0 & 0 & 0 & 2\,q_{4} & 2\,q_{3} & -2\ell_{33} & 2\,q_{1} & 0 & 0 & 0 & 0 & 0\\ 0 & 0 & 0 & 0 & 0 & 0 & 2\,q_{4} & 2\,q_{3} & 2\,q_{2} & -2\ell_{11} & 2\,q_{3} & 2\,q_{2} & 2\,q_{1} & 0 & 0\\ 0 & 0 & 0 & 0 & 0 & 0 & 0 & 0 & 0 & 2\,q_{4} & -2\ell_{22} & 2\,q_{2} & 2\,q_{1} & 0 & 0\\ 0 & 0 & 0 & 0 & 0 & 0 & 0 & 0 & 0 & 2\,q_{4} & 2\,q_{3} & -2\ell_{33} & 2\,q_{1} & 0 & 0\\ 2\,q_{1} & 0 & 0 & 0 & 0 & 0 & 0 & 0 & 0 & 2\,q_{4} & 2\,q_{3} & 2\,q_{2} & -2\ell_{11} & 2\,q_{3} & 2\,q_{2}\\ 2\,q_{1} & 0 & 0 & 0 & 0 & 0 & 0 & 0 & 0 & 0 & 0 & 0 & 2\,q_{4} & -2\ell_{22} & 2\,q_{2}\\ 2\,q_{1} & 0 & 0 & 0 & 0 & 0 & 0 & 0 & 0 & 0 & 0 & 0 & 2\,q_{4} & 2\,q_{3} & -2\ell_{33}\end{array}\right)}\, ,
\end{equation*}
where
\begin{equation}
    \ell_{11} = q_{4}+2q_{3}+2q_{2}+q_{1}\, , \ell_{22}=q_{4}+q_{2}+q_{1}\, ,\ell_{33}= q_{4}+q_{3}+q_{1}\, ,
\end{equation}

Let us finally observe that the Laplace matrix $\mathbf{L}^{(3)}$ of the undirected $3$-hyperring is given by
\begin{equation*}
   \mathbf{L}^{(3)}= \left(\begin{array}{ccccccccccccccc} -6 & 1 & 1 & 1 & 0 & 0 & 0 & 0 & 0 & 0 & 0 & 0 & 1 & 1 & 1\\ 1 & -3 & 1 & 1 & 0 & 0 & 0 & 0 & 0 & 0 & 0 & 0 & 0 & 0 & 0\\ 1 & 1 & -3 & 1 & 0 & 0 & 0 & 0 & 0 & 0 & 0 & 0 & 0 & 0 & 0\\ 1 & 1 & 1 & -6 & 1 & 1 & 1 & 0 & 0 & 0 & 0 & 0 & 0 & 0 & 0\\ 0 & 0 & 0 & 1 & -3 & 1 & 1 & 0 & 0 & 0 & 0 & 0 & 0 & 0 & 0\\ 0 & 0 & 0 & 1 & 1 & -3 & 1 & 0 & 0 & 0 & 0 & 0 & 0 & 0 & 0\\ 0 & 0 & 0 & 1 & 1 & 1 & -6 & 1 & 1 & 1 & 0 & 0 & 0 & 0 & 0\\ 0 & 0 & 0 & 0 & 0 & 0 & 1 & -3 & 1 & 1 & 0 & 0 & 0 & 0 & 0\\ 0 & 0 & 0 & 0 & 0 & 0 & 1 & 1 & -3 & 1 & 0 & 0 & 0 & 0 & 0\\ 0 & 0 & 0 & 0 & 0 & 0 & 1 & 1 & 1 & -6 & 1 & 1 & 1 & 0 & 0\\ 0 & 0 & 0 & 0 & 0 & 0 & 0 & 0 & 0 & 1 & -3 & 1 & 1 & 0 & 0\\ 0 & 0 & 0 & 0 & 0 & 0 & 0 & 0 & 0 & 1 & 1 & -3 & 1 & 0 & 0\\ 1 & 0 & 0 & 0 & 0 & 0 & 0 & 0 & 0 & 1 & 1 & 1 & -6 & 1 & 1\\ 1 & 0 & 0 & 0 & 0 & 0 & 0 & 0 & 0 & 0 & 0 & 0 & 1 & -3 & 1\\ 1 & 0 & 0 & 0 & 0 & 0 & 0 & 0 & 0 & 0 & 0 & 0 & 1 & 1 & -3 \end{array}\right)\, ,
\end{equation*}
and thus $\mathbf{M}^{(3,3)}(q_1,q_2,q_3,q_4)=\hat{\mathbf{M}}^{(3,3)}+\check{\mathbf{M}}^{(3,3)}$ coincide with $\mathbf{L}^{(3)}$ if $q_1=q_2=q_3=q_4=1/4$.

\section{Construction of the $m$-directed random hypergraph}
\label{sec:mdirrandHG}

The goal of this section is to present more details about the construction of the $1$-directed random hypergraph used in the main text.

The model is inspired by the one presented in~\cite{PhysRevE.106.064310}, indeed we fix a number $N$ of nodes and a number $Q$ of hyperedges. Each hyperedge size $e_i$ is drawn from a binomial distribution with parameters $(N,p)$ for some fixed $p\in (0,1)$, then $|e_i|$ nodes are uniformly randomly drawn without repetitions from the set $[1,N]$ and added to the hyperedge $e_i$. The process is repeated for $i=1,\dots, Q$, by paying attention to not reproduce twice the same hypergraph, i.e., the same combination of nodes. Eventually $m$ nodes are uniformly selected from each hyperedge and they are assigned to the hyperedge head, and thus the remaining $q_i=|e_i|-m$ nodes form the tail. In the initial construction we have also to pay attention that each hyperedge size is large enough to contain $m$ head nodes and at least one tail node. To completely remove pairwise interactions, we also added the constraint that $|e_i|\geq 3$.

\end{document}